\documentclass[aps,prl,showpacs,twocolumn,superscriptaddress]{revtex4-1}

\usepackage{graphicx}
\usepackage{color}
\usepackage{pgf}
\usepackage{enumerate}
\usepackage{amsmath}
\usepackage{amssymb}
\usepackage{stmaryrd}
\usepackage{multirow}
\usepackage{float}
\usepackage{booktabs}
\usepackage{physics}
\usepackage{url}
\usepackage{soul} 
\usepackage{braket}
\usepackage[caption=false]{subfig}
\usepackage[draft]{pdfcomment}

\def\beq{\begin{equation}}
\def\eeq{\end{equation}}	 
\usepackage[normalem]{ulem}

\newcommand*{\matminus}{%
  \leavevmode
  \hphantom{0}%
  \llap{%
    \settowidth{\dimen0 }{$0$}%
    \resizebox{1.1\dimen0 }{\height}{$-$}%
  }%
}

\begin{document}

\title{Mitigating the Sign Problem Through Basis Rotations}

\author{Ryan Levy}
\affiliation{Institute for Condensed Matter Theory and IQUIST and Department of Physics, University of Illinois at Urbana-Champaign, IL 61801, USA} 

\author{Bryan K.  Clark}
\affiliation{Institute for Condensed Matter Theory and IQUIST and Department of Physics, University of Illinois at Urbana-Champaign, IL 61801, USA} 
\begin{abstract}
Quantum Monte Carlo simulations of quantum many body systems are plagued by the Fermion sign problem. The computational complexity of simulating Fermions scales exponentially in the projection time $\beta$ and system size. The sign problem is basis dependent and an improved basis, for fixed errors, lead to exponentially quicker simulations.  We show how to use sign-free quantum Monte Carlo simulations to optimize over the choice of basis on large two-dimensional systems.  
We numerically illustrate these techniques decreasing the `badness' of the sign problem by optimizing over single-particle basis rotations on one and two-dimensional Hubbard systems.  We find a generic rotation which improves the average sign of the Hubbard model for a wide range of $U$ and densities for $L \times 4$ systems.  In one example improvement, the average sign (and hence simulation cost at fixed accuracy) for the $16\times 4$ Hubbard model at $U/t=4$ and $n=0.75$ increases by $\exp\left[8.64(6)\beta\right]$. For typical projection times of $\beta\gtrapprox 100$, this accelerates such simulation by many orders of magnitude.  

\end{abstract}
 
\maketitle

The power of quantum computers and the difficulty of simulating quantum many-body ground states stem from similar sources:  non-trivial entanglement and paths of interfering signs.  Low entanglement states can be solved using the density matrix renormalization group (DMRG) \cite{schollwock2011density,WhiteDMRG}.  Paths of interfering signs are the source of the fermion sign problem in quantum Monte Carlo (QMC) simulations.  Systems, such as bosonic Hamiltonians, which have only positive paths and hence no sign problem, can be simulated efficiently.  For some classes of systems with `naive' sign-problems, approaches have been found which also allow for efficient simulations.   One common technique involves finding a sign-problem free basis in which to perform the simulation \cite{Nakamura1998,Okunishi2014,Honecker2016,Alet2016,Wessel2018}.  Examples where this has also been successful include Hubbard models at half-filling \cite{Wang2015,Li2018} and employing a Majorana representation \cite{Li2015}.  Note that while there is always a sign-problem free basis (i.e. the eigenstate basis),  it is often as (or more) difficult to find this basis as doing the simulation.  Even in cases where the sign problem can't be removed, the choice of basis can affect the `badness' of the sign problem \cite{kolodrubetz2012effect}.
Interestingly, there has not yet been a significant amount of work in automatically searching for a good basis to help mitigate or remove the sign problem.  It is this problem that we approach in this Letter. 

\begin{figure}[t]

\noindent \centering{}\includegraphics[width=0.66\columnwidth]{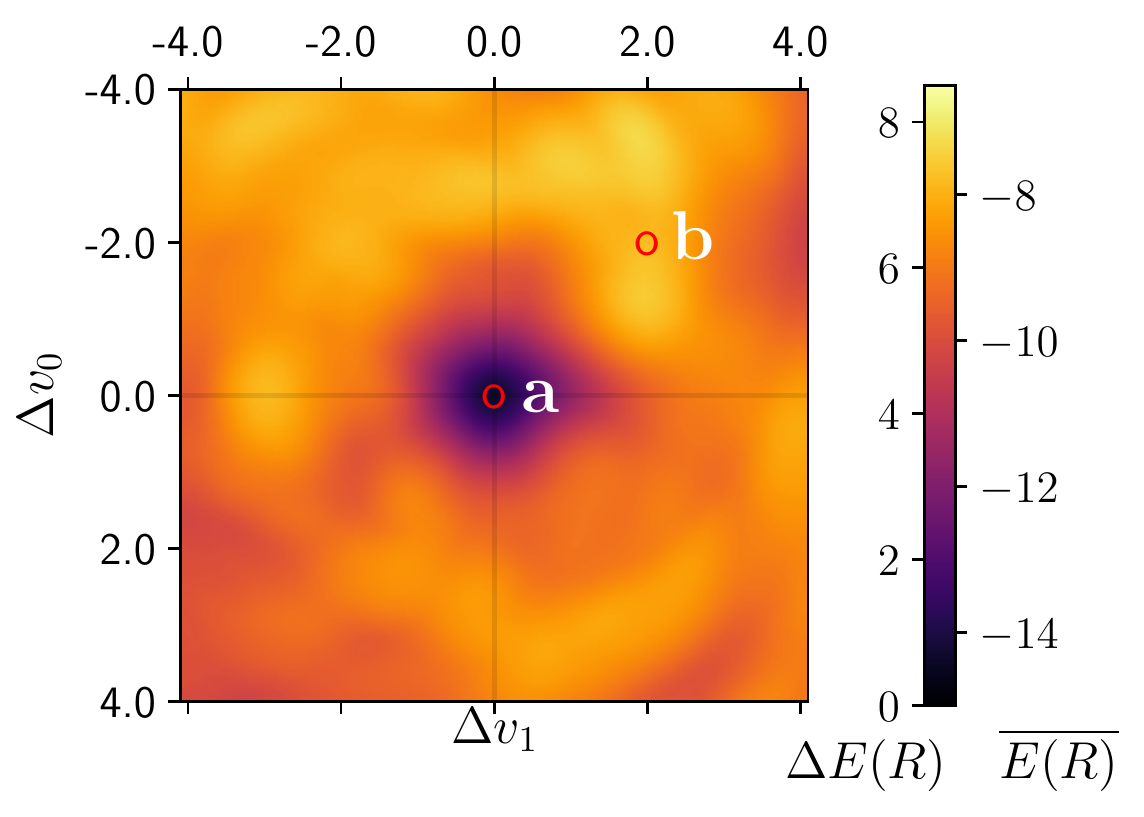}
 \includegraphics[width=0.315\columnwidth]{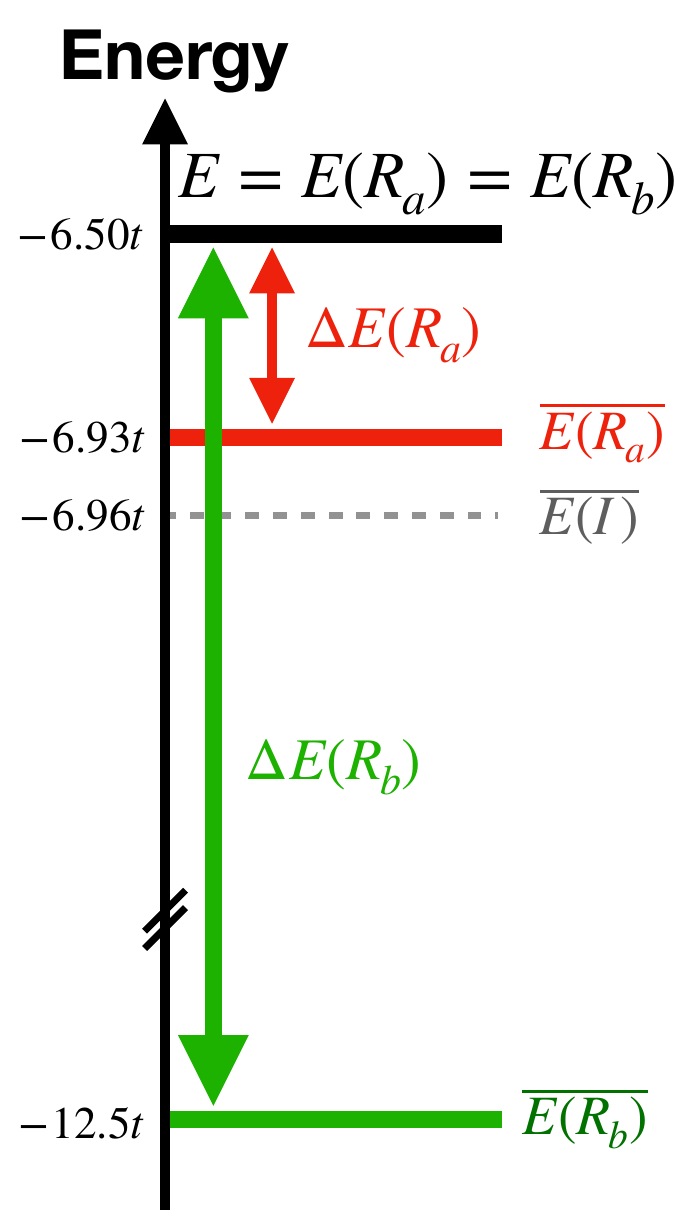}

\caption{\textit{Left:}  $\Delta E(R)$ and $\overline{E(R)}$ for a 8 site Hubbard chain at $n=0.5$ and $U/t=1$ in the plane of the first two parameters; the final plot is interpolated. Dark/light colors indicate mitigation/increased sign problem respectively.  Colorbar shows both $\Delta E$ (which measures the `badness' of the sign  problem but cannot be computed efficiently) as well as $\overline{E(R)}$ which can be efficiently computed for large systems.
\textit{Right:} Depiction (not to scale) of the energies of two particular rotations $R_a$ and $R_b$. Despite having the same energy $E$ when there's a sign problem, by optimizing for the largest sign-free $\overline{E(R)}$ we minimize the gap $\Delta E(R)$ which exponentially improves the average sign $\langle s \rangle$ by $\exp[\beta (\overline{E(R)}-\overline{E(I)})]\approx \exp[0.03\beta]$ for projection time $\beta$. }
\label{Fig:L3Signs} 
\end{figure}

Our approach  is as follows: given a quantification of the sign problem, search over a class of unitary rotations of the Hamiltonian minimizing the `badness' of the sign problem.  The key question, then, is the development of an algorithm to efficiently accomplish this.  As a proof of principle we will consider hole-doped Hubbard models finding a rotation that decreases the `badness' of the sign problem for $L \times 4$ lattices over a wide range of $U/t$.

\section{Quantification of Sign Problem}
Let  $H$ be the Hamiltonian represented in a standard basis (i.e. real space) and $H(R) \equiv RHR^\dagger$ be the Hamiltonian rotated by the unitary rotation $R$. Our choice for quantifying the sign problem of $H(R)$ has two goals: (1) The objective function should be proportional to the cost of QMC on $H(R)$.  (2) We can measure (and hence optimize) this objective function efficiently on large bulk two-dimensional systems. 

The natural objective function to maximize is the relative variance of the average sign  
\begin{equation}
    \langle s \rangle \equiv \frac{1}{N}\sum_i s_i
\end{equation} where $s_i$ is the sign for each Monte Carlo sample $i$ and $N$ is the total number of Monte Carlo samples.  While the efficiency of different observables vary, their values
$\langle O \rangle = \sum_i s_i O_i/\langle s\rangle$ all involve the average sign.  Here $O_i$ is the observable measured on Monte Carlo sample $i$.
Given the variance of $\langle s \rangle$ is $O(1)$, minimizing the relative variance comes down to maximizing $\langle s \rangle.$  Because our interest will be in the limit of a large QMC projection time $\beta$ our focus will be in optimizing the component of $\langle s \rangle  = A \overline{s}^\beta$  which causes $\langle s \rangle$ to decay exponentially with $\beta$ - i.e. $\overline{s}$.

Naively, though, this appears to be difficult to optimize as computing $\langle s \rangle$ itself has a sign problem.  It turns out, though, that locally optimizing $\langle s(R) \rangle$ over a class of unitary rotations $R$ is significantly easier then computing it.  In particular, $\overline{s(R)}=\exp(-\Delta E(R))$  \cite{kolodrubetz2012effect,spencer2012sign,ceperley1981stochastic} where $\Delta E(R) \equiv E(R) - \overline{E(R)}$ is the gap between the ground state $E(R)$ of the fermionic Hamiltonian $H(R)$ and the ground state $\overline{E(R)}$ of an effective bosonic version of the Hamiltonian $|H(R)|$ \cite{PhysRevB.41.9301,HATANO1992,Troyer2005a} defined as 
\begin{align}
    |H(R)|_{ij} &  = -H(R)_{ij} \textrm{ if } i \neq j \textrm{ and } H(R)_{ij}>0 \nonumber \\
     &  = H(R)_{ij} \textrm{ otherwise }
\end{align}
with 
$\overline{E(R)}= \langle \overline{ \Psi} | \overline{H(R)} | \overline{\Psi}\rangle$

where $|\overline{\Psi}\rangle$ is the ground state of $\overline{H(R)}$.
Note this definition ensures that the propagator used in QMC, $|G|\equiv 1 - \tau |H|$ (where $\tau$  is a small constant) contains strictly positive matrix elements.  While $E(R)$  is hard to compute (there is a sign problem in evaluating it), its value is independent of the choice of the unitary rotation - i.e. $E(R)=E$.  On the other hand $\overline{E(R)}$ is sign-free and hence computable by Monte Carlo on large systems.  Maximizing $\overline{s(R)}$ then comes down to maximizing $\overline{E(R)}$.

Note this metric is closely related to, but not identical to having a ground state with only positive amplitudes or a Hamiltonian with no positive off-diagonal matrix  elements.  For example, the Heisenberg model on a bipartite lattice has no sign problem (i.e. no exponentially large relative variance) in QMC but does have positive off-diagonal Hamiltonian matrix elements and a non-positive ground state (although both these can be removed by applying the Marshall sign rule \cite{Marshall1955}).  $\Delta E(I)$ where $I$ is the identity rotation is correctly zero in this case.

\section{Unitary Rotation}
Here we describe the parameterization of our unitaries. While a unitary rotation exists which removes the sign problem (i.e. the eigenstate basis), it is often harder to find then the ground state itself.  Instead, we focus on unitary rotations which have simple representations.  Two such classes are unitary rotations from shallow-depth quantum circuits and basis rotations on the set of single-particle orbitals used to represent the Hamiltonian: i.e. given a Hamiltonian  written as $H=\sum_{ij} t_{ij} c_i^\dagger c_j + \sum_{ijkl} V_{ijkl} c_i^\dagger c_j^\dagger c_k c_l$ we can write $b_j  \equiv \sum_k U_{jk} c_k$ where $U$ is a $N \times N$ unitary matrix for a system of $N$ sites.  $U$ can be parameterized in various ways including as the eigenvalues of an orthogonal matrix $H$  or as $ U = e^{A}$
where $A$ is a skew-symmetric  matrix, i.e.  $A=-A^T$.
In our numerical examples, we focus on single particle basis rotations although the approach we describe works also with quantum circuits.

\section{Optimization}
Our goal now is to maximize $\overline{E(R)}$.  In this section we demonstrate a series of techniques to optimize this quantity which we benchmark on single particle basis rotations of the Hubbard model under periodic boundary conditions,
\begin{align}
    H=-t\sum_{\sigma\langle i,j\rangle}c_{j\sigma}^{\dagger}c_{i\sigma}+h.c.+U\sum_{i}n_{i\uparrow}n_{i\downarrow}
\end{align}
with $\langle i,j \rangle$ denoting nearest neighbors. 

We optimize $\overline{E(R)}$ via 3 separate techniques. First, we optimize $\overline{E(R)}$ via exact diagonalization using finite differences. When diagonalization is too costly, we turn to the second technique, projector quantum Monte Carlo \cite{Nandini1990,Ceperly1980} (PQMC).  PQMC is a form of QMC which propagates walkers for a projection time 
$\beta$ giving the energy 
\begin{equation}
    \overline{E(R;\beta)} = \frac{\langle \Psi_\textrm{init} | \overline{H(R)} \exp[-\beta \overline{H(R)}] |\Psi_\textrm{init} \rangle}{\langle \Psi_\textrm{init} | \exp[-\beta \overline{H(R)}] |\Psi_\textrm{init}\rangle}
\end{equation}  where walkers are sampled from an initial wave-function $|\Psi_\textrm{init}\rangle$.  In PQMC, by taking large $\beta$, we can compute $\overline{E(R)} \equiv \lim_{\beta \rightarrow \infty} \overline{E(R;\beta)}$ in polynomial time; derivatives can be computed using finite differences giving an algorithm which scales linearly in the number of parameters and otherwise is similar in cost to a sign-free QMC simulation of the same rotated Hamiltonian (i.e. polynomial in system size). 

Parameters are then updated  using the optimization scheme described in \cite{luo2019backflow} inspired by \cite{JieSandvik2007}. For each of the unique entries $v_i$ of our unitary parameterization, the next parameter is determined by
\begin{align}
    v_{i+1} = v_i + \alpha \gamma \frac{\left| \partial\overline{E}/\partial v_i\right|}{ \partial\overline{E}/\partial v_i},
\end{align}
where $\alpha$ is a random number chosen between 0 and 1 and $\gamma$ controls the size of the random step.  

\begin{figure}
        \centering
        \subfloat[\label{Fig:optimization}]{
            \includegraphics[width=\columnwidth]{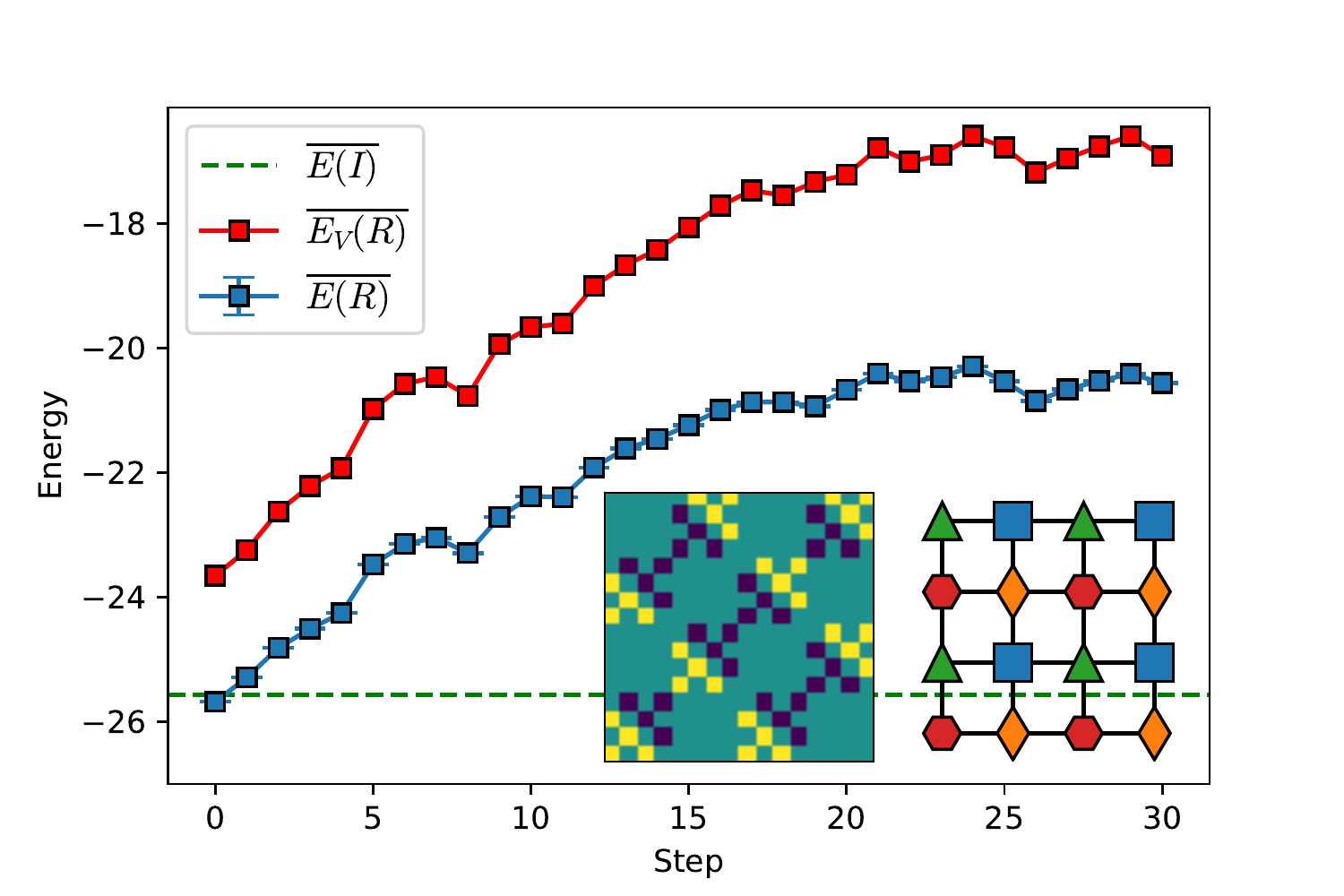}
        }\\
        \subfloat[\label{Fig:optimization16x4}]{%
            \centering 
            \includegraphics[width=\columnwidth]{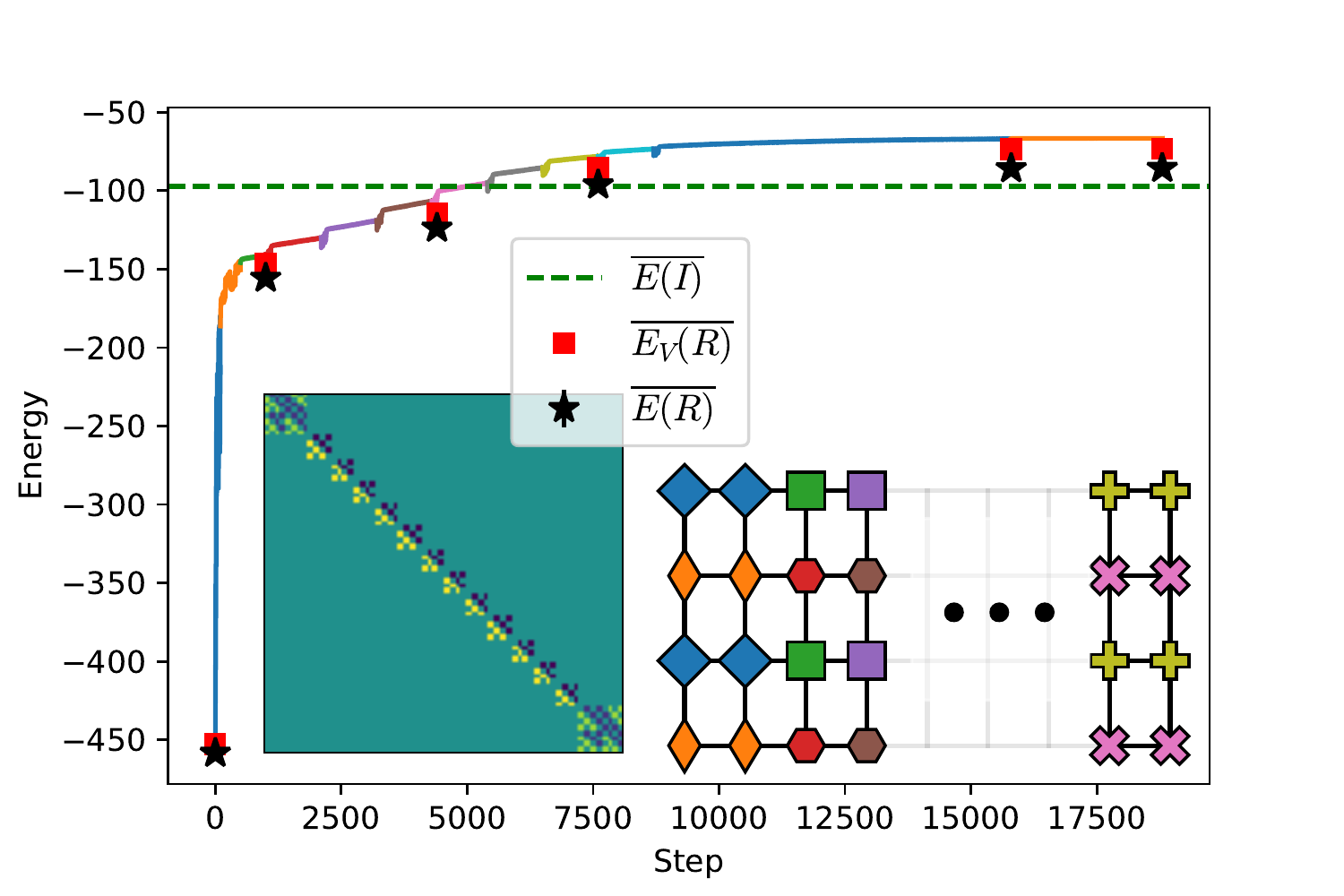}
            \label{fig:mean and std of net34}
        }
        \caption{ (a) Projector optimization of $4\times 4$ Hubbard model with $\gamma=0.01$ showing $\overline{E(R)}$ (blue) and $\overline{E_V(R)}$ (red). 
        (b) Optimization of a $16\times 4$ Hubbard model showing both noisy values of  $\overline{E_V(R)}$ used in optimization (colored lines) as well as accurate values of   $\overline{E_V(R)}$ (red squares) and $\overline{E(R)}$ (stars).  Each color indicates a change in hyperparameters of optimization (see supplementary). 
        \textit{Left insets:} Rounded optimal unitary rotations at the end of optimization, in a matrix representation.
        \textit{Right insets:} Lattices illustrating the new basis. The rotated basis is made up of linear combinations of 2 or 4 operators, marked by their same color and shape. 
        } 
        \label{Fig:OptPlots}
    \end{figure}

Although the cost of the direct PQMC approach is polynomial it can still be expensive especially since this rotation can result in a quadratic number of terms per (matrix) row (which is a significant increase from the linear number of terms in the Hubbard model but standard for QMC on materials \cite{doi:10.1063/1.3193710}).  To overcome this expense, we develop and benchmark a further approximation to our algorithm which significantly accelerates the optimization. Instead of directly optimizing  $\overline{E(R)}$, we optimize 
\begin{equation}
    \overline{E_V(R)} \equiv \langle \Psi_\textrm{u(R)} | \overline{H(R)} | \Psi_\textrm{u(R)} \rangle
\end{equation} where $\Psi_\textrm{u(R)} \equiv 1/2^{N}$ is the uniform wave-function \textit{in the basis $R$}.  This is a strict upper bound in energy to $\overline{H(R)}$  and is equivalent to $\overline{E(R;\beta=0)}$ in the PQMC when the initial wave-function is chosen as $|\Psi_\textrm{u(R)}\rangle$.  This approximation turns out to be reasonable as the variational energy $\overline{E_V(R)}$ tracks the ground state energy $\overline{E(R)}$ and so pushes the parameters in the correct direction (see fig.~\ref{Fig:OptPlots}).  This approximation can be made even faster by choosing a relatively fixed few random states ($O(10)$) and using these states to optimize $\overline{E_V(R)}$. We further utilize the Jax library \cite{jax2018github} and implement the derivative $d\overline{E_V(R)}/dR$ using automatic differentiation.
Because these derivatives (and the energy) are computed from only $O(10)$ Monte Carlo samples, the statistical error is large, but the correlations introduced by using the same configurations better controls the path generated by the gradients.

\subsection{Results}

\subsubsection{1D}
We perform initial tests optimizing one-dimensional Hubbard chains for $N=\{3,4,8\}$. We evaluate $\Delta E(R)$ via exact diagonalization and perform derivative free optimization using Nelder–Mead optimization \cite{nlopt,nelder1965simplex}.  In these runs, $R$ is parameterized by the eigenvectors of a real Hermitian matrix.  Our optimization completely removes the sign-problem at $N=3$ at 2/3 doping and $N=4$ Hubbard model at half filling for all values of $U$ (see supplementary material for the specific matrices); \textit{a posteri}, we discovered these rotations are known in the CTQMC literature \cite{Shinaoka2015}.  
 Interestingly at $N=4$ (see supplementary), our optimization procedure finds a basis which does not make all the off-diagonal signs of $|G|$ positive despite still removing the sign problem  (although by hand $G$ can be made positive). We also include an example for $N=8$ where we improved but were unable to completely remove the sign problem (see fig.~\ref{Fig:L3Signs}).  
\subsubsection{2D}
Next we consider a variety of width-4 Hubbard models.
First we implement projector optimization for the $n=0.5$, $U/t=1$, $4 \times 4$ Hubbard model. We parameterize $R$ as the matrix exponential of a real skew-symmetric matrix and start from a random unitary which has a bad sign problem (worse than not rotating at all).

Fig.~\ref{Fig:optimization} shows the improvement in $\overline{E(R)}$ during optimization. By rounding the final unitary and removing noise, we obtain a sparse unitary rotation which rotates each site operator into linear combinations of 4-operator plaquettes (see inset of fig.~\ref{Fig:optimization}). The rotated operators don't map to their counterpart in the original basis, i.e. operator $b_{(i,j)}$ doesn't contain $c_{(i,j)}$. We find that $\Delta E(R)=0.137$ and achieve an increase in the average sign which goes as $\exp[6.27\beta]$ at large $\beta$.

\begin{figure}[ht]
\noindent \centering{}\includegraphics[width=\columnwidth]{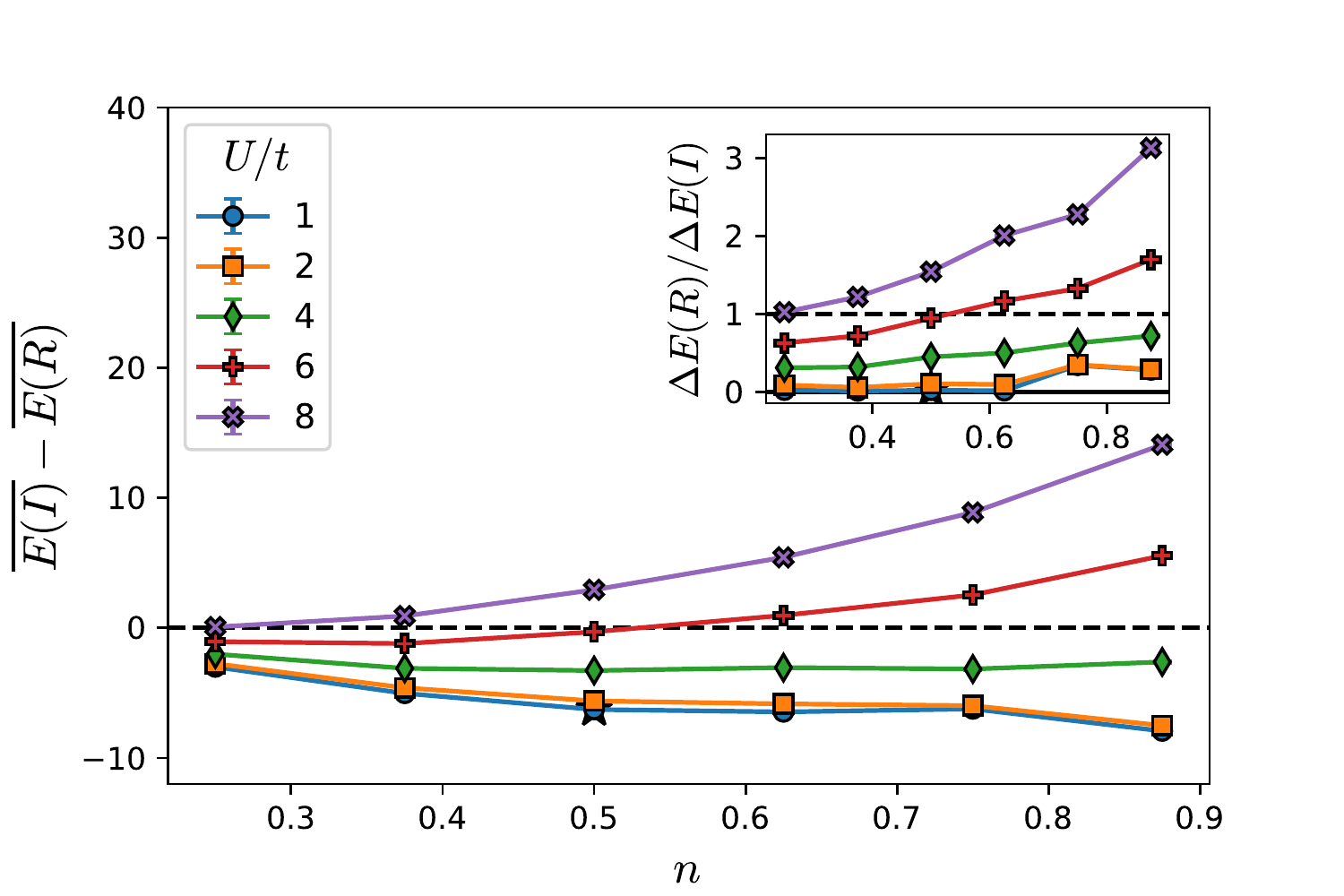}

\caption{Comparing the results of projector optimization of a $4\times4$ Hubbard Model at $U/t=1$ and $n=0.5$ doping (star) to no basis rotation ($I$) at different $U$ and doping values. PQMC is done in continuous time with a resampling rate of  $T=0.1$ to a projection time of $\beta=20$.
\textit{Inset:} Relative $\Delta E(R)/\Delta E(I)$ between the optimal rotation and no basis rotation ($R=I$).}
\label{Fig:dopeOptimization} 
\end{figure}

We test our optimized rotation for a range of doping $n$ and interaction strengths $U$ in fig.~\ref{Fig:dopeOptimization}. While only optimized for a single $U$ and $n$, we find this rotation mitigates the sign problem over the real-space basis for the $4 \times 4$ Hubbard model for all $0.125<n<1.0$ when $U/t\leq4$.
The ability for a single optimized rotation to be applicable to a larger regime of phase space shows the versatility of our approach. 

We then turn to width-4 cylinders with periodic boundary conditions. We perform optimization on the $16 \times 4$ Hubbard model with $n=0.5$ and $U/t=2$, using automatic differentiation of $\overline{E_V(R)}$ and varying step sizes (see supplementary), shown in fig.~\ref{Fig:optimization16x4}. By rounding and removing noise of the final unitary we obtain a sparse rotation which improves the sign by $\exp[12.86(5)\beta]$. We observe this rotation also works for a variety of doping and $U$ values, e.g. $n=0.5$ $U/t=6$ and $n=0.75$ $U/t=4$ have improvements of the average sign by $\exp[3.78(6)\beta]$ and $\exp[8.64(6)\beta]$ respectively.

Upon analyzing the unitary, we find a clear structure that can be used for any width-4 cylinder. The single particle orbitals in the bulk of the lattice become a linear combination over each column of width four, leading to 4 creation/annihilation operators (with implicit spin indices) 
\begin{equation}
   \begin{pmatrix}b_{(n,4)}\\
b_{(n,3)}\\
b_{(n,2)}\\
b_{(n,1)}
\end{pmatrix}^{(\dagger)}=\dfrac{1}{\sqrt{2}}\begin{pmatrix}
0 & \matminus1 & 0 & \matminus 1\\
1 & 0 & \matminus1 & 0\\
0 & 1 & 0 & \matminus1\\
1 & 0 & 1 & 0
\end{pmatrix}\begin{pmatrix}c_{(n,4)}\\
c_{(n,3)}\\
c_{(n,2)}\\
c_{(n,1)}
\end{pmatrix}^{(\dagger)} 
\label{eq:colRot}
\end{equation}

At the left and right edges of the system, despite periodic boundaries, two consecutive columns are rotated as
\begin{equation}
    \left(\begin{smallmatrix}b_{(2,4)}\\
\vdots\\
b_{(2,1)}\\
b_{(1,4)}\\
\vdots\\
b_{(1,1)}
\end{smallmatrix}\right)^{(\dagger)}
=\dfrac{1}{2}\left(\begin{smallmatrix}
0 & \matminus1 & 0 & \matminus1 & 0 & 1 & 0 & 1\\
1 & 0 & \matminus1 & 0 & \matminus1 & 0 & 1 & 0\\
0 & 1 & 0 & \matminus1 & 0 & \matminus1 & 0 & 1\\
1 & 0 & 1 & 0 & \matminus1 & 0 & \matminus1 & 0\\
0 & \matminus1 & 0 & \matminus1 & 0 & \matminus1 & 0 & \matminus1\\
1 & 0 & \matminus1 & 0 & 1 & 0 & \matminus1 & 0\\
0 & 1 & 0 & \matminus1 & 0 & 1 & 0 & \matminus1\\
1 & 0 & 1 & 0 & 1 & 0 & 1 & 0
\end{smallmatrix}\right)
 \left(\begin{smallmatrix}c_{(2,4)}\\
\vdots\\
c_{(2,1)}\\
c_{(1,4)}\\
\vdots\\
c_{(1,1)}
\end{smallmatrix}\right)^{(\dagger)}.
\label{eq:edgeRot}
\end{equation}
We illustrate this rotation in the inset of fig.~\ref{Fig:optimization16x4}. Similar to the structure of the $4\times4$ rotation, the new operators are linear combinations of nearby sites but lack the original site with respect to the old basis. However, the bulk is made of 2-site pairs of sites rather than 4 site plaquettes, and the edge plaquettes are not spaced apart in the $\hat{x}$ direction. 

Testing this for a larger system, we observe an average sign reduction over not rotating on a $32\times 4$ cylinder at $n=0.5$ doping of $\exp[22.5(4)\beta]$ and $\exp[16.0(3)\beta]$ for $U/t=4$ and $U/t=6$ respectively. These energy differences $\overline{E(I)} -\overline{E(R)}$ per site at $U=4$ are increasing with respect to system size (from $L=8$ to $L=32$), suggesting in the thermodynamic limit that our rotation provides a better lower bound to the true fermionic energy than identity rotation (details can be found in the supplementary). 

Finally, we can take the edge rotation found in eq.~\ref{eq:edgeRot} and apply it both to the $4\times 4$ and $8\times8$ lattices. For the $4\times 4$ case, we find that the rotation improves the sign more not than rotating, but not as well as our optimized $4\times 4$ unitary, i.e. $\exp[5.40\beta]$ compared to $\exp[6.27\beta]$ for $n=1$ $U/t=1$ respectively. When using this rotation to tile an $8\times8$ unitary, we find that we improve the sign only for $U/t\lesssim 1$, e.g. at $U/t=1$ we find an improvement of $\exp[-3.0(1)\beta]$ and $\exp[-3.9(1)\beta]$ for $n=0.5$ and $n=0.875$ respectively.

In addition to the above results, we include in the supplementary an optimization of a 6x6 lattice to show that the original PQMC optimization of $\overline{E(R)}$ scales beyond what is accessible by exact diagonalization. We also include a study on $4 \times 4$ systems of how the rotation affects the average sign in PQMC with respect to $\beta$ and the benefit of introducing annihilation \cite{doi:10.1063/1.3193710,Kalos2000} on top of the rotation.  

\subsection{Conclusion}
We have introduced a new approach for finding a basis which reduces the sign problem in bulk systems of arbitrary dimensions.
For fixed accuracy and projection time, this improves the calculation speed of a ground-state QMC by many orders of magnitude; a similar approach will also improve finite-temperature calculations.   Our method uses QMC to maximize the energy of the ``bosonic'' version of the rotated Hamiltonian $|H(R)|$; this maximization corresponds to maximizing the average sign of a QMC calculation in that rotated basis.  In many ways, our sign-problem mitigation approach is similar in flavor to variational Monte Carlo\cite{Umrigar2007,Neuscamman2012,kochkov2018variational,otis2019complementary,zaletel2019isometric,haghshenas2019canonicalization,carleo2017solving,kochkov2018variational} and many of the tricks from that community will be applicable here.   Although we used PQMC in this Letter, any QMC which works in an orthogonal basis (world-line Monte Carlo, path integral ground state, etc.) have the same sign problem and are equally improved by the rotation.  
In addition, we developed  a variational optimization approach which instead optimizes the variational energy of the uniform state $|\Psi_{u(R)}\rangle$ with respect to $|H(R)|$ over a small number of configurations, for a significant speedup and at polynomial scaling.  This heuristic works because of the empirical observation that this variational energy and the projector energy track each other closely up to some (slowly varying) fixed offset. 

To illustrate the method, we optimized the single-particle basis of $4\times 4$ and $16\times 4$ Hubbard model at $n=0.5$ doping and $U/t=1$ and $U/t=2$ respectively. 
This resulted in an exponential increase in average sign, by approximately $\exp\left[ 6.278\beta \right]$ and $\exp\left[ 12.86(5)\beta \right]$ leading to significant speedups for a  typical $\beta\in [10,100]$.  The rotations we found are sparse, generalize to a wider range of $U$ and $n$, and the latter rotation generalizes to all width-4 cylinders.

A uniting theme among these optimized rotations is their structure of rotating into linear combinations of a small number of nearby sites. The new basis in both cases seem relatively insensitive to both doping and $U/t$, with each $b_{(i,j)}$ not including the matching $c_{(i,j)}$.  In addition, a 4-operator plaquette is apparent in both geometries, although in slightly different forms. 
Other techniques beyond QMC simulations may also benefit from this insight, although its unclear if it extends beyond the Hubbard model itself. The automatic optimization of our method should be useful for other sign-plagued models to potentially turn up other sparse bases.

\begin{acknowledgments} \textit{Acknowledgments:---} We acknowledge Luiz Santos for helping to initiate this work trying to find a sign-problem mitigating unitary, early collaboration on this project, and useful discussions and insights related to this work.  
BKC thanks David Ceperley with whom he collaborated on a related unpublished project as a graduate student that attempted to mitigate the sign problem by minimizing the sum of the badly-signed off-diagonal matrix elements. BKC also thanks Shiwei Zhang, Miles Stoudenmire, and Giuseppe Carleo for conversations.   RL thanks Shivesh Pathak for useful conversations and Yubo Yang for visualization assistance. 
We acknowledge support from the Department of Energy (DOE) Award No. DE-SC0020165.
This project is part of the Blue
Waters sustained-petascale computing project, which is supported by the National Science Foundation (awards OCI-0725070 and ACI-1238993) and the State of Illinois. Blue Waters is a joint effort of the University of Illinois at Urbana-Champaign and its National Center for Supercomputing Applications. 
This work also made use of the Illinois Campus Cluster, a computing resource that is operated by the Illinois Campus Cluster Program (ICCP) in conjunction with the National Center for Supercomputing Applications (NCSA) and which is supported by funds from the University of Illinois at Urbana-Champaign. 

\textit{Note:} During the final stages of this work, a number of related works appeared on the arXiv \cite{Torlai2019}, \cite{Hangleiter}, and \cite{Klassen2019}. 
\end{acknowledgments}
\bibliography{refs}

\begin{thebibliography}{39}%
\makeatletter
\providecommand \@ifxundefined [1]{%
 \@ifx{#1\undefined}
}%
\providecommand \@ifnum [1]{%
 \ifnum #1\expandafter \@firstoftwo
 \else \expandafter \@secondoftwo
 \fi
}%
\providecommand \@ifx [1]{%
 \ifx #1\expandafter \@firstoftwo
 \else \expandafter \@secondoftwo
 \fi
}%
\providecommand \natexlab [1]{#1}%
\providecommand \enquote  [1]{``#1''}%
\providecommand \bibnamefont  [1]{#1}%
\providecommand \bibfnamefont [1]{#1}%
\providecommand \citenamefont [1]{#1}%
\providecommand \href@noop [0]{\@secondoftwo}%
\providecommand \href [0]{\begingroup \@sanitize@url \@href}%
\providecommand \@href[1]{\@@startlink{#1}\@@href}%
\providecommand \@@href[1]{\endgroup#1\@@endlink}%
\providecommand \@sanitize@url [0]{\catcode `\\12\catcode `\$12\catcode
  `\&12\catcode `\#12\catcode `\^12\catcode `\_12\catcode `\%12\relax}%
\providecommand \@@startlink[1]{}%
\providecommand \@@endlink[0]{}%
\providecommand \url  [0]{\begingroup\@sanitize@url \@url }%
\providecommand \@url [1]{\endgroup\@href {#1}{\urlprefix }}%
\providecommand \urlprefix  [0]{URL }%
\providecommand \Eprint [0]{\href }%
\providecommand \doibase [0]{http://dx.doi.org/}%
\providecommand \selectlanguage [0]{\@gobble}%
\providecommand \bibinfo  [0]{\@secondoftwo}%
\providecommand \bibfield  [0]{\@secondoftwo}%
\providecommand \translation [1]{[#1]}%
\providecommand \BibitemOpen [0]{}%
\providecommand \bibitemStop [0]{}%
\providecommand \bibitemNoStop [0]{.\EOS\space}%
\providecommand \EOS [0]{\spacefactor3000\relax}%
\providecommand \BibitemShut  [1]{\csname bibitem#1\endcsname}%
\let\auto@bib@innerbib\@empty
\bibitem [{\citenamefont {Schollw{\"o}ck}(2011)}]{schollwock2011density}%
  \BibitemOpen
  \bibfield  {author} {\bibinfo {author} {\bibfnamefont {U.}~\bibnamefont
  {Schollw{\"o}ck}},\ }\href@noop {} {\bibfield  {journal} {\bibinfo  {journal}
  {Annals of Physics}\ }\textbf {\bibinfo {volume} {326}},\ \bibinfo {pages}
  {96} (\bibinfo {year} {2011})}\BibitemShut {NoStop}%
\bibitem [{\citenamefont {White}(1992)}]{WhiteDMRG}%
  \BibitemOpen
  \bibfield  {author} {\bibinfo {author} {\bibfnamefont {S.~R.}\ \bibnamefont
  {White}},\ }\href {\doibase 10.1103/PhysRevLett.69.2863} {\bibfield
  {journal} {\bibinfo  {journal} {Phys. Rev. Lett.}\ }\textbf {\bibinfo
  {volume} {69}},\ \bibinfo {pages} {2863} (\bibinfo {year}
  {1992})}\BibitemShut {NoStop}%
\bibitem [{\citenamefont {Nakamura}(1998)}]{Nakamura1998}%
  \BibitemOpen
  \bibfield  {author} {\bibinfo {author} {\bibfnamefont {T.}~\bibnamefont
  {Nakamura}},\ }\href {\doibase 10.1103/PhysRevB.57.R3197} {\bibfield
  {journal} {\bibinfo  {journal} {Phys. Rev. B}\ }\textbf {\bibinfo {volume}
  {57}},\ \bibinfo {pages} {R3197} (\bibinfo {year} {1998})},\ \Eprint
  {http://arxiv.org/abs/9707019} {9707019} \BibitemShut {NoStop}%
\bibitem [{\citenamefont {Okunishi}\ and\ \citenamefont
  {Harada}(2014)}]{Okunishi2014}%
  \BibitemOpen
  \bibfield  {author} {\bibinfo {author} {\bibfnamefont {K.}~\bibnamefont
  {Okunishi}}\ and\ \bibinfo {author} {\bibfnamefont {K.}~\bibnamefont
  {Harada}},\ }\href {\doibase 10.1103/PhysRevB.89.134422} {\bibfield
  {journal} {\bibinfo  {journal} {Phys. Rev. B}\ }\textbf {\bibinfo {volume}
  {89}},\ \bibinfo {pages} {134422} (\bibinfo {year} {2014})}\BibitemShut
  {NoStop}%
\bibitem [{\citenamefont {Honecker}\ \emph {et~al.}(2016)\citenamefont
  {Honecker}, \citenamefont {Wessel}, \citenamefont {Kerkdyk}, \citenamefont
  {Pruschke}, \citenamefont {Mila},\ and\ \citenamefont
  {Normand}}]{Honecker2016}%
  \BibitemOpen
  \bibfield  {author} {\bibinfo {author} {\bibfnamefont {A.}~\bibnamefont
  {Honecker}}, \bibinfo {author} {\bibfnamefont {S.}~\bibnamefont {Wessel}},
  \bibinfo {author} {\bibfnamefont {R.}~\bibnamefont {Kerkdyk}}, \bibinfo
  {author} {\bibfnamefont {T.}~\bibnamefont {Pruschke}}, \bibinfo {author}
  {\bibfnamefont {F.}~\bibnamefont {Mila}}, \ and\ \bibinfo {author}
  {\bibfnamefont {B.}~\bibnamefont {Normand}},\ }\href {\doibase
  10.1103/PhysRevB.93.054408} {\bibfield  {journal} {\bibinfo  {journal} {Phys.
  Rev. B}\ }\textbf {\bibinfo {volume} {93}},\ \bibinfo {pages} {054408}
  (\bibinfo {year} {2016})}\BibitemShut {NoStop}%
\bibitem [{\citenamefont {Alet}\ \emph {et~al.}(2016)\citenamefont {Alet},
  \citenamefont {Damle},\ and\ \citenamefont {Pujari}}]{Alet2016}%
  \BibitemOpen
  \bibfield  {author} {\bibinfo {author} {\bibfnamefont {F.}~\bibnamefont
  {Alet}}, \bibinfo {author} {\bibfnamefont {K.}~\bibnamefont {Damle}}, \ and\
  \bibinfo {author} {\bibfnamefont {S.}~\bibnamefont {Pujari}},\ }\href
  {\doibase 10.1103/PhysRevLett.117.197203} {\bibfield  {journal} {\bibinfo
  {journal} {Phys. Rev. Lett.}\ }\textbf {\bibinfo {volume} {117}},\ \bibinfo
  {pages} {197203} (\bibinfo {year} {2016})}\BibitemShut {NoStop}%
\bibitem [{\citenamefont {Wessel}\ \emph {et~al.}(2018)\citenamefont {Wessel},
  \citenamefont {Niesen}, \citenamefont {Stapmanns}, \citenamefont {Normand},
  \citenamefont {Mila}, \citenamefont {Corboz},\ and\ \citenamefont
  {Honecker}}]{Wessel2018}%
  \BibitemOpen
  \bibfield  {author} {\bibinfo {author} {\bibfnamefont {S.}~\bibnamefont
  {Wessel}}, \bibinfo {author} {\bibfnamefont {I.}~\bibnamefont {Niesen}},
  \bibinfo {author} {\bibfnamefont {J.}~\bibnamefont {Stapmanns}}, \bibinfo
  {author} {\bibfnamefont {B.}~\bibnamefont {Normand}}, \bibinfo {author}
  {\bibfnamefont {F.}~\bibnamefont {Mila}}, \bibinfo {author} {\bibfnamefont
  {P.}~\bibnamefont {Corboz}}, \ and\ \bibinfo {author} {\bibfnamefont
  {A.}~\bibnamefont {Honecker}},\ }\href {\doibase 10.1103/PhysRevB.98.174432}
  {\bibfield  {journal} {\bibinfo  {journal} {Phys. Rev. B}\ }\textbf {\bibinfo
  {volume} {98}},\ \bibinfo {pages} {174432} (\bibinfo {year}
  {2018})}\BibitemShut {NoStop}%
\bibitem [{\citenamefont {Wang}\ \emph {et~al.}(2015)\citenamefont {Wang},
  \citenamefont {Liu}, \citenamefont {Iazzi}, \citenamefont {Troyer},\ and\
  \citenamefont {Harcos}}]{Wang2015}%
  \BibitemOpen
  \bibfield  {author} {\bibinfo {author} {\bibfnamefont {L.}~\bibnamefont
  {Wang}}, \bibinfo {author} {\bibfnamefont {Y.-H.}\ \bibnamefont {Liu}},
  \bibinfo {author} {\bibfnamefont {M.}~\bibnamefont {Iazzi}}, \bibinfo
  {author} {\bibfnamefont {M.}~\bibnamefont {Troyer}}, \ and\ \bibinfo {author}
  {\bibfnamefont {G.}~\bibnamefont {Harcos}},\ }\href {\doibase
  10.1103/PhysRevLett.115.250601} {\bibfield  {journal} {\bibinfo  {journal}
  {Phys. Rev. Lett.}\ }\textbf {\bibinfo {volume} {115}},\ \bibinfo {pages}
  {250601} (\bibinfo {year} {2015})}\BibitemShut {NoStop}%
\bibitem [{\citenamefont {Li}\ and\ \citenamefont {Yao}(2019)}]{Li2018}%
  \BibitemOpen
  \bibfield  {author} {\bibinfo {author} {\bibfnamefont {Z.-X.}\ \bibnamefont
  {Li}}\ and\ \bibinfo {author} {\bibfnamefont {H.}~\bibnamefont {Yao}},\
  }\href {\doibase 10.1146/annurev-conmatphys-033117-054307} {\bibfield
  {journal} {\bibinfo  {journal} {Annual Review of Condensed Matter Physics}\
  }\textbf {\bibinfo {volume} {10}},\ \bibinfo {pages} {337} (\bibinfo {year}
  {2019})},\ \Eprint {http://arxiv.org/abs/1805.08219} {1805.08219}
  \BibitemShut {NoStop}%
\bibitem [{\citenamefont {Li}\ \emph {et~al.}(2015)\citenamefont {Li},
  \citenamefont {Jiang},\ and\ \citenamefont {Yao}}]{Li2015}%
  \BibitemOpen
  \bibfield  {author} {\bibinfo {author} {\bibfnamefont {Z.-X.}\ \bibnamefont
  {Li}}, \bibinfo {author} {\bibfnamefont {Y.-F.}\ \bibnamefont {Jiang}}, \
  and\ \bibinfo {author} {\bibfnamefont {H.}~\bibnamefont {Yao}},\ }\href
  {\doibase 10.1103/PhysRevB.91.241117} {\bibfield  {journal} {\bibinfo
  {journal} {Phys. Rev. B}\ }\textbf {\bibinfo {volume} {91}},\ \bibinfo
  {pages} {241117(R)} (\bibinfo {year} {2015})}\BibitemShut {NoStop}%
\bibitem [{\citenamefont {Kolodrubetz}\ \emph {et~al.}(2013)\citenamefont
  {Kolodrubetz}, \citenamefont {Spencer}, \citenamefont {Clark},\ and\
  \citenamefont {Foulkes}}]{kolodrubetz2012effect}%
  \BibitemOpen
  \bibfield  {author} {\bibinfo {author} {\bibfnamefont {M.~H.}\ \bibnamefont
  {Kolodrubetz}}, \bibinfo {author} {\bibfnamefont {J.~S.}\ \bibnamefont
  {Spencer}}, \bibinfo {author} {\bibfnamefont {B.~K.}\ \bibnamefont {Clark}},
  \ and\ \bibinfo {author} {\bibfnamefont {W.~M.~C.}\ \bibnamefont {Foulkes}},\
  }\href@noop {} {\bibfield  {journal} {\bibinfo  {journal} {The Journal of
  chemical physics}\ }\textbf {\bibinfo {volume} {138}},\ \bibinfo {pages}
  {024110} (\bibinfo {year} {2013})}\BibitemShut {NoStop}%
\bibitem [{\citenamefont {Spencer}\ \emph {et~al.}(2012)\citenamefont
  {Spencer}, \citenamefont {Blunt},\ and\ \citenamefont
  {Foulkes}}]{spencer2012sign}%
  \BibitemOpen
  \bibfield  {author} {\bibinfo {author} {\bibfnamefont {J.}~\bibnamefont
  {Spencer}}, \bibinfo {author} {\bibfnamefont {N.}~\bibnamefont {Blunt}}, \
  and\ \bibinfo {author} {\bibfnamefont {W.}~\bibnamefont {Foulkes}},\
  }\href@noop {} {\bibfield  {journal} {\bibinfo  {journal} {The Journal of
  chemical physics}\ }\textbf {\bibinfo {volume} {136}},\ \bibinfo {pages}
  {054110} (\bibinfo {year} {2012})}\BibitemShut {NoStop}%
\bibitem [{\citenamefont {Ceperley}(1981)}]{ceperley1981stochastic}%
  \BibitemOpen
  \bibfield  {author} {\bibinfo {author} {\bibfnamefont {D.}~\bibnamefont
  {Ceperley}},\ }in\ \href@noop {} {\emph {\bibinfo {booktitle} {Recent
  Progress in Many-Body Theories}}}\ (\bibinfo  {publisher} {Springer},\
  \bibinfo {year} {1981})\ pp.\ \bibinfo {pages} {262--269}\BibitemShut
  {NoStop}%
\bibitem [{\citenamefont {Loh}\ \emph {et~al.}(1990)\citenamefont {Loh},
  \citenamefont {Gubernatis}, \citenamefont {Scalettar}, \citenamefont {White},
  \citenamefont {Scalapino},\ and\ \citenamefont {Sugar}}]{PhysRevB.41.9301}%
  \BibitemOpen
  \bibfield  {author} {\bibinfo {author} {\bibfnamefont {E.~Y.}\ \bibnamefont
  {Loh}}, \bibinfo {author} {\bibfnamefont {J.~E.}\ \bibnamefont {Gubernatis}},
  \bibinfo {author} {\bibfnamefont {R.~T.}\ \bibnamefont {Scalettar}}, \bibinfo
  {author} {\bibfnamefont {S.~R.}\ \bibnamefont {White}}, \bibinfo {author}
  {\bibfnamefont {D.~J.}\ \bibnamefont {Scalapino}}, \ and\ \bibinfo {author}
  {\bibfnamefont {R.~L.}\ \bibnamefont {Sugar}},\ }\href {\doibase
  10.1103/PhysRevB.41.9301} {\bibfield  {journal} {\bibinfo  {journal} {Phys.
  Rev. B}\ }\textbf {\bibinfo {volume} {41}},\ \bibinfo {pages} {9301}
  (\bibinfo {year} {1990})}\BibitemShut {NoStop}%
\bibitem [{\citenamefont {Hatano}\ and\ \citenamefont
  {Suzuki}(1992)}]{HATANO1992}%
  \BibitemOpen
  \bibfield  {author} {\bibinfo {author} {\bibfnamefont {N.}~\bibnamefont
  {Hatano}}\ and\ \bibinfo {author} {\bibfnamefont {M.}~\bibnamefont
  {Suzuki}},\ }\href {\doibase https://doi.org/10.1016/0375-9601(92)91006-D}
  {\bibfield  {journal} {\bibinfo  {journal} {Physics Letters A}\ }\textbf
  {\bibinfo {volume} {163}},\ \bibinfo {pages} {246 } (\bibinfo {year}
  {1992})}\BibitemShut {NoStop}%
\bibitem [{\citenamefont {Troyer}\ and\ \citenamefont
  {Wiese}(2005)}]{Troyer2005a}%
  \BibitemOpen
  \bibfield  {author} {\bibinfo {author} {\bibfnamefont {M.}~\bibnamefont
  {Troyer}}\ and\ \bibinfo {author} {\bibfnamefont {U.-J.}\ \bibnamefont
  {Wiese}},\ }\href {\doibase 10.1103/PhysRevLett.94.170201} {\bibfield
  {journal} {\bibinfo  {journal} {Phys. Rev. Lett.}\ }\textbf {\bibinfo
  {volume} {94}},\ \bibinfo {pages} {170201} (\bibinfo {year}
  {2005})}\BibitemShut {NoStop}%
\bibitem [{\citenamefont {Marshall}\ and\ \citenamefont
  {Peierls}(1955)}]{Marshall1955}%
  \BibitemOpen
  \bibfield  {author} {\bibinfo {author} {\bibfnamefont {W.}~\bibnamefont
  {Marshall}}\ and\ \bibinfo {author} {\bibfnamefont {R.~E.}\ \bibnamefont
  {Peierls}},\ }\href {\doibase 10.1098/rspa.1955.0200} {\bibfield  {journal}
  {\bibinfo  {journal} {Proc. R. Soc. London. A.}\ }\textbf {\bibinfo {volume}
  {232}},\ \bibinfo {pages} {48} (\bibinfo {year} {1955})}\BibitemShut
  {NoStop}%
\bibitem [{\citenamefont {Trivedi}\ and\ \citenamefont
  {Ceperley}(1990)}]{Nandini1990}%
  \BibitemOpen
  \bibfield  {author} {\bibinfo {author} {\bibfnamefont {N.}~\bibnamefont
  {Trivedi}}\ and\ \bibinfo {author} {\bibfnamefont {D.~M.}\ \bibnamefont
  {Ceperley}},\ }\href {\doibase 10.1103/PhysRevB.41.4552} {\bibfield
  {journal} {\bibinfo  {journal} {Phys. Rev. B}\ }\textbf {\bibinfo {volume}
  {41}},\ \bibinfo {pages} {4552} (\bibinfo {year} {1990})}\BibitemShut
  {NoStop}%
\bibitem [{\citenamefont {Ceperley}\ and\ \citenamefont
  {Alder}(1980)}]{Ceperly1980}%
  \BibitemOpen
  \bibfield  {author} {\bibinfo {author} {\bibfnamefont {D.~M.}\ \bibnamefont
  {Ceperley}}\ and\ \bibinfo {author} {\bibfnamefont {B.~J.}\ \bibnamefont
  {Alder}},\ }\href {\doibase 10.1103/PhysRevLett.45.566} {\bibfield  {journal}
  {\bibinfo  {journal} {Phys. Rev. Lett.}\ }\textbf {\bibinfo {volume} {45}},\
  \bibinfo {pages} {566} (\bibinfo {year} {1980})}\BibitemShut {NoStop}%
\bibitem [{\citenamefont {Luo}\ and\ \citenamefont
  {Clark}(2019)}]{luo2019backflow}%
  \BibitemOpen
  \bibfield  {author} {\bibinfo {author} {\bibfnamefont {D.}~\bibnamefont
  {Luo}}\ and\ \bibinfo {author} {\bibfnamefont {B.~K.}\ \bibnamefont
  {Clark}},\ }\href@noop {} {\bibfield  {journal} {\bibinfo  {journal}
  {Physical Review Letters}\ }\textbf {\bibinfo {volume} {122}},\ \bibinfo
  {pages} {226401} (\bibinfo {year} {2019})}\BibitemShut {NoStop}%
\bibitem [{\citenamefont {Lou}\ and\ \citenamefont
  {Sandvik}(2007)}]{JieSandvik2007}%
  \BibitemOpen
  \bibfield  {author} {\bibinfo {author} {\bibfnamefont {J.}~\bibnamefont
  {Lou}}\ and\ \bibinfo {author} {\bibfnamefont {A.~W.}\ \bibnamefont
  {Sandvik}},\ }\href {\doibase 10.1103/PhysRevB.76.104432} {\bibfield
  {journal} {\bibinfo  {journal} {Phys. Rev. B}\ }\textbf {\bibinfo {volume}
  {76}},\ \bibinfo {pages} {104432} (\bibinfo {year} {2007})}\BibitemShut
  {NoStop}%
\bibitem [{\citenamefont {Booth}\ \emph {et~al.}(2009)\citenamefont {Booth},
  \citenamefont {Thom},\ and\ \citenamefont {Alavi}}]{doi:10.1063/1.3193710}%
  \BibitemOpen
  \bibfield  {author} {\bibinfo {author} {\bibfnamefont {G.~H.}\ \bibnamefont
  {Booth}}, \bibinfo {author} {\bibfnamefont {A.~J.~W.}\ \bibnamefont {Thom}},
  \ and\ \bibinfo {author} {\bibfnamefont {A.}~\bibnamefont {Alavi}},\ }\href
  {\doibase 10.1063/1.3193710} {\bibfield  {journal} {\bibinfo  {journal} {The
  Journal of Chemical Physics}\ }\textbf {\bibinfo {volume} {131}},\ \bibinfo
  {pages} {054106} (\bibinfo {year} {2009})}\BibitemShut {NoStop}%
\bibitem [{\citenamefont {Bradbury}\ \emph {et~al.}(2018)\citenamefont
  {Bradbury}, \citenamefont {Frostig}, \citenamefont {Hawkins}, \citenamefont
  {Johnson}, \citenamefont {Leary}, \citenamefont {Maclaurin}, \citenamefont
  {Necula}, \citenamefont {Paszke}, \citenamefont {Vander{P}las}, \citenamefont
  {Wanderman-{M}ilne},\ and\ \citenamefont {Zhang}}]{jax2018github}%
  \BibitemOpen
  \bibfield  {author} {\bibinfo {author} {\bibfnamefont {J.}~\bibnamefont
  {Bradbury}}, \bibinfo {author} {\bibfnamefont {R.}~\bibnamefont {Frostig}},
  \bibinfo {author} {\bibfnamefont {P.}~\bibnamefont {Hawkins}}, \bibinfo
  {author} {\bibfnamefont {M.~J.}\ \bibnamefont {Johnson}}, \bibinfo {author}
  {\bibfnamefont {C.}~\bibnamefont {Leary}}, \bibinfo {author} {\bibfnamefont
  {D.}~\bibnamefont {Maclaurin}}, \bibinfo {author} {\bibfnamefont
  {G.}~\bibnamefont {Necula}}, \bibinfo {author} {\bibfnamefont
  {A.}~\bibnamefont {Paszke}}, \bibinfo {author} {\bibfnamefont
  {J.}~\bibnamefont {Vander{P}las}}, \bibinfo {author} {\bibfnamefont
  {S.}~\bibnamefont {Wanderman-{M}ilne}}, \ and\ \bibinfo {author}
  {\bibfnamefont {Q.}~\bibnamefont {Zhang}},\ }\href
  {http://github.com/google/jax} {\enquote {\bibinfo {title} {{JAX}: composable
  transformations of {P}ython+{N}um{P}y programs},}\ } (\bibinfo {year}
  {2018})\BibitemShut {NoStop}%
\bibitem [{\citenamefont {Johnson}()}]{nlopt}%
  \BibitemOpen
  \bibfield  {author} {\bibinfo {author} {\bibfnamefont {S.~G.}\ \bibnamefont
  {Johnson}},\ }\href {http://github.com/stevengj/nlopt} {\enquote {\bibinfo
  {title} {{ The NLopt nonlinear-optimization package}},}\ }\BibitemShut
  {NoStop}%
\bibitem [{\citenamefont {Nelder}\ and\ \citenamefont
  {Mead}(1965)}]{nelder1965simplex}%
  \BibitemOpen
  \bibfield  {author} {\bibinfo {author} {\bibfnamefont {J.~A.}\ \bibnamefont
  {Nelder}}\ and\ \bibinfo {author} {\bibfnamefont {R.}~\bibnamefont {Mead}},\
  }\href {\doibase 10.1093/comjnl/7.4.308} {\bibfield  {journal} {\bibinfo
  {journal} {The computer journal}\ }\textbf {\bibinfo {volume} {7}},\ \bibinfo
  {pages} {308} (\bibinfo {year} {1965})}\BibitemShut {NoStop}%
\bibitem [{\citenamefont {Shinaoka}\ \emph {et~al.}(2015)\citenamefont
  {Shinaoka}, \citenamefont {Nomura}, \citenamefont {Biermann}, \citenamefont
  {Troyer},\ and\ \citenamefont {Werner}}]{Shinaoka2015}%
  \BibitemOpen
  \bibfield  {author} {\bibinfo {author} {\bibfnamefont {H.}~\bibnamefont
  {Shinaoka}}, \bibinfo {author} {\bibfnamefont {Y.}~\bibnamefont {Nomura}},
  \bibinfo {author} {\bibfnamefont {S.}~\bibnamefont {Biermann}}, \bibinfo
  {author} {\bibfnamefont {M.}~\bibnamefont {Troyer}}, \ and\ \bibinfo {author}
  {\bibfnamefont {P.}~\bibnamefont {Werner}},\ }\href {\doibase
  10.1103/PhysRevB.92.195126} {\bibfield  {journal} {\bibinfo  {journal} {Phys.
  Rev. B}\ }\textbf {\bibinfo {volume} {92}},\ \bibinfo {pages} {195126}
  (\bibinfo {year} {2015})}\BibitemShut {NoStop}%
\bibitem [{\citenamefont {Kalos}\ and\ \citenamefont
  {Pederiva}(2000)}]{Kalos2000}%
  \BibitemOpen
  \bibfield  {author} {\bibinfo {author} {\bibfnamefont {M.~H.}\ \bibnamefont
  {Kalos}}\ and\ \bibinfo {author} {\bibfnamefont {F.}~\bibnamefont
  {Pederiva}},\ }\href {\doibase 10.1103/PhysRevLett.85.3547} {\bibfield
  {journal} {\bibinfo  {journal} {Phys. Rev. Lett.}\ }\textbf {\bibinfo
  {volume} {85}},\ \bibinfo {pages} {3547} (\bibinfo {year}
  {2000})}\BibitemShut {NoStop}%
\bibitem [{\citenamefont {Umrigar}\ \emph {et~al.}(2007)\citenamefont
  {Umrigar}, \citenamefont {Toulouse}, \citenamefont {Filippi}, \citenamefont
  {Sorella},\ and\ \citenamefont {Hennig}}]{Umrigar2007}%
  \BibitemOpen
  \bibfield  {author} {\bibinfo {author} {\bibfnamefont {C.~J.}\ \bibnamefont
  {Umrigar}}, \bibinfo {author} {\bibfnamefont {J.}~\bibnamefont {Toulouse}},
  \bibinfo {author} {\bibfnamefont {C.}~\bibnamefont {Filippi}}, \bibinfo
  {author} {\bibfnamefont {S.}~\bibnamefont {Sorella}}, \ and\ \bibinfo
  {author} {\bibfnamefont {R.~G.}\ \bibnamefont {Hennig}},\ }\href {\doibase
  10.1103/PhysRevLett.98.110201} {\bibfield  {journal} {\bibinfo  {journal}
  {Phys. Rev. Lett.}\ }\textbf {\bibinfo {volume} {98}},\ \bibinfo {pages}
  {110201} (\bibinfo {year} {2007})}\BibitemShut {NoStop}%
\bibitem [{\citenamefont {Neuscamman}\ \emph {et~al.}(2012)\citenamefont
  {Neuscamman}, \citenamefont {Umrigar},\ and\ \citenamefont
  {Chan}}]{Neuscamman2012}%
  \BibitemOpen
  \bibfield  {author} {\bibinfo {author} {\bibfnamefont {E.}~\bibnamefont
  {Neuscamman}}, \bibinfo {author} {\bibfnamefont {C.~J.}\ \bibnamefont
  {Umrigar}}, \ and\ \bibinfo {author} {\bibfnamefont {G.~K.-L.}\ \bibnamefont
  {Chan}},\ }\href {\doibase 10.1103/PhysRevB.85.045103} {\bibfield  {journal}
  {\bibinfo  {journal} {Phys. Rev. B}\ }\textbf {\bibinfo {volume} {85}},\
  \bibinfo {pages} {045103} (\bibinfo {year} {2012})}\BibitemShut {NoStop}%
\bibitem [{\citenamefont {Kochkov}\ and\ \citenamefont
  {Clark}(2018)}]{kochkov2018variational}%
  \BibitemOpen
  \bibfield  {author} {\bibinfo {author} {\bibfnamefont {D.}~\bibnamefont
  {Kochkov}}\ and\ \bibinfo {author} {\bibfnamefont {B.~K.}\ \bibnamefont
  {Clark}},\ }\href@noop {} {\bibfield  {journal} {\bibinfo  {journal} {arXiv
  preprint arXiv:1811.12423}\ } (\bibinfo {year} {2018})}\BibitemShut {NoStop}%
\bibitem [{\citenamefont {Otis}\ and\ \citenamefont
  {Neuscamman}(2019)}]{otis2019complementary}%
  \BibitemOpen
  \bibfield  {author} {\bibinfo {author} {\bibfnamefont {L.}~\bibnamefont
  {Otis}}\ and\ \bibinfo {author} {\bibfnamefont {E.}~\bibnamefont
  {Neuscamman}},\ }\href@noop {} {\bibfield  {journal} {\bibinfo  {journal}
  {Physical Chemistry Chemical Physics}\ } (\bibinfo {year}
  {2019})}\BibitemShut {NoStop}%
\bibitem [{\citenamefont {Zaletel}\ and\ \citenamefont
  {Pollmann}(2019)}]{zaletel2019isometric}%
  \BibitemOpen
  \bibfield  {author} {\bibinfo {author} {\bibfnamefont {M.~P.}\ \bibnamefont
  {Zaletel}}\ and\ \bibinfo {author} {\bibfnamefont {F.}~\bibnamefont
  {Pollmann}},\ }\href@noop {} {\bibfield  {journal} {\bibinfo  {journal}
  {arXiv preprint arXiv:1902.05100}\ } (\bibinfo {year} {2019})}\BibitemShut
  {NoStop}%
\bibitem [{\citenamefont {Haghshenas}\ \emph {et~al.}(2019)\citenamefont
  {Haghshenas}, \citenamefont {O'Rourke},\ and\ \citenamefont
  {Chan}}]{haghshenas2019canonicalization}%
  \BibitemOpen
  \bibfield  {author} {\bibinfo {author} {\bibfnamefont {R.}~\bibnamefont
  {Haghshenas}}, \bibinfo {author} {\bibfnamefont {M.~J.}\ \bibnamefont
  {O'Rourke}}, \ and\ \bibinfo {author} {\bibfnamefont {G.~K.}\ \bibnamefont
  {Chan}},\ }\href@noop {} {\bibfield  {journal} {\bibinfo  {journal} {arXiv
  preprint arXiv:1903.03843}\ } (\bibinfo {year} {2019})}\BibitemShut {NoStop}%
\bibitem [{\citenamefont {Carleo}\ and\ \citenamefont
  {Troyer}(2017)}]{carleo2017solving}%
  \BibitemOpen
  \bibfield  {author} {\bibinfo {author} {\bibfnamefont {G.}~\bibnamefont
  {Carleo}}\ and\ \bibinfo {author} {\bibfnamefont {M.}~\bibnamefont
  {Troyer}},\ }\href@noop {} {\bibfield  {journal} {\bibinfo  {journal}
  {Science}\ }\textbf {\bibinfo {volume} {355}},\ \bibinfo {pages} {602}
  (\bibinfo {year} {2017})}\BibitemShut {NoStop}%
\bibitem [{\citenamefont {Torlai}\ \emph {et~al.}(2020)\citenamefont {Torlai},
  \citenamefont {Carrasquilla}, \citenamefont {Fishman}, \citenamefont
  {Melko},\ and\ \citenamefont {Fisher}}]{Torlai2019}%
  \BibitemOpen
  \bibfield  {author} {\bibinfo {author} {\bibfnamefont {G.}~\bibnamefont
  {Torlai}}, \bibinfo {author} {\bibfnamefont {J.}~\bibnamefont
  {Carrasquilla}}, \bibinfo {author} {\bibfnamefont {M.~T.}\ \bibnamefont
  {Fishman}}, \bibinfo {author} {\bibfnamefont {R.~G.}\ \bibnamefont {Melko}},
  \ and\ \bibinfo {author} {\bibfnamefont {M.~P.~A.}\ \bibnamefont {Fisher}},\
  }\href {http://arxiv.org/abs/1906.04654} {\bibfield  {journal} {\bibinfo
  {journal} {Phys. Rev. Research}\ }\textbf {\bibinfo {volume} {2}},\ \bibinfo
  {pages} {032060(R)} (\bibinfo {year} {2020})},\ \Eprint
  {http://arxiv.org/abs/1906.04654} {arXiv:1906.04654} \BibitemShut {NoStop}%
\bibitem [{\citenamefont {Hangleiter}\ \emph {et~al.}(2020)\citenamefont
  {Hangleiter}, \citenamefont {Roth}, \citenamefont {Nagaj},\ and\
  \citenamefont {Eisert}}]{Hangleiter}%
  \BibitemOpen
  \bibfield  {author} {\bibinfo {author} {\bibfnamefont {D.}~\bibnamefont
  {Hangleiter}}, \bibinfo {author} {\bibfnamefont {I.}~\bibnamefont {Roth}},
  \bibinfo {author} {\bibfnamefont {D.}~\bibnamefont {Nagaj}}, \ and\ \bibinfo
  {author} {\bibfnamefont {J.}~\bibnamefont {Eisert}},\ }\href {\doibase
  10.1126/sciadv.abb8341} {\bibfield  {journal} {\bibinfo  {journal} {Science
  Advances}\ }\textbf {\bibinfo {volume} {6}} (\bibinfo {year} {2020}),\
  10.1126/sciadv.abb8341}\BibitemShut {NoStop}%
\bibitem [{\citenamefont {Klassen}\ \emph {et~al.}(2020)\citenamefont
  {Klassen}, \citenamefont {Marvian}, \citenamefont {Piddock}, \citenamefont
  {Ioannou}, \citenamefont {Hen},\ and\ \citenamefont {Terhal}}]{Klassen2019}%
  \BibitemOpen
  \bibfield  {author} {\bibinfo {author} {\bibfnamefont {J.}~\bibnamefont
  {Klassen}}, \bibinfo {author} {\bibfnamefont {M.}~\bibnamefont {Marvian}},
  \bibinfo {author} {\bibfnamefont {S.}~\bibnamefont {Piddock}}, \bibinfo
  {author} {\bibfnamefont {M.}~\bibnamefont {Ioannou}}, \bibinfo {author}
  {\bibfnamefont {I.}~\bibnamefont {Hen}}, \ and\ \bibinfo {author}
  {\bibfnamefont {B.}~\bibnamefont {Terhal}},\ }\href {\doibase
  10.1137/19M1287511} {\bibfield  {journal} {\bibinfo  {journal} {SIAM Journal
  on Computing}\ }\textbf {\bibinfo {volume} {49}},\ \bibinfo {pages} {1332}
  (\bibinfo {year} {2020})},\ \Eprint {http://arxiv.org/abs/1906.08800}
  {arXiv:1906.08800} \BibitemShut {NoStop}%
\bibitem [{\citenamefont {Dagotto}\ \emph {et~al.}(1992)\citenamefont
  {Dagotto}, \citenamefont {Moreo}, \citenamefont {Ortolani}, \citenamefont
  {Poilblanc},\ and\ \citenamefont {Riera}}]{Dagotto1992}%
  \BibitemOpen
  \bibfield  {author} {\bibinfo {author} {\bibfnamefont {E.}~\bibnamefont
  {Dagotto}}, \bibinfo {author} {\bibfnamefont {A.}~\bibnamefont {Moreo}},
  \bibinfo {author} {\bibfnamefont {F.}~\bibnamefont {Ortolani}}, \bibinfo
  {author} {\bibfnamefont {D.}~\bibnamefont {Poilblanc}}, \ and\ \bibinfo
  {author} {\bibfnamefont {J.}~\bibnamefont {Riera}},\ }\href {\doibase
  10.1103/PhysRevB.45.10741} {\bibfield  {journal} {\bibinfo  {journal} {Phys.
  Rev. B}\ }\textbf {\bibinfo {volume} {45}},\ \bibinfo {pages} {10741}
  (\bibinfo {year} {1992})}\BibitemShut {NoStop}%
\bibitem [{\citenamefont {LeBlanc}\ \emph {et~al.}(2015)\citenamefont
  {LeBlanc}, \citenamefont {Antipov}, \citenamefont {Becca}, \citenamefont
  {Bulik}, \citenamefont {Chan}, \citenamefont {Chung}, \citenamefont {Deng},
  \citenamefont {Ferrero}, \citenamefont {Henderson}, \citenamefont
  {Jim{\'{e}}nez-Hoyos}, \citenamefont {Kozik}, \citenamefont {Liu},
  \citenamefont {Millis}, \citenamefont {Prokof'ev}, \citenamefont {Qin},
  \citenamefont {Scuseria}, \citenamefont {Shi}, \citenamefont {Svistunov},
  \citenamefont {Tocchio}, \citenamefont {Tupitsyn}, \citenamefont {White},
  \citenamefont {Zhang}, \citenamefont {Zheng}, \citenamefont {Zhu},\ and\
  \citenamefont {Gull}}]{LeBlanc2015}%
  \BibitemOpen
  \bibfield  {author} {\bibinfo {author} {\bibfnamefont {J.~P.~F.}\
  \bibnamefont {LeBlanc}}, \bibinfo {author} {\bibfnamefont {A.~E.}\
  \bibnamefont {Antipov}}, \bibinfo {author} {\bibfnamefont {F.}~\bibnamefont
  {Becca}}, \bibinfo {author} {\bibfnamefont {I.~W.}\ \bibnamefont {Bulik}},
  \bibinfo {author} {\bibfnamefont {G.~K.-L.}\ \bibnamefont {Chan}}, \bibinfo
  {author} {\bibfnamefont {C.-M.}\ \bibnamefont {Chung}}, \bibinfo {author}
  {\bibfnamefont {Y.}~\bibnamefont {Deng}}, \bibinfo {author} {\bibfnamefont
  {M.}~\bibnamefont {Ferrero}}, \bibinfo {author} {\bibfnamefont {T.~M.}\
  \bibnamefont {Henderson}}, \bibinfo {author} {\bibfnamefont {C.~A.}\
  \bibnamefont {Jim{\'{e}}nez-Hoyos}}, \bibinfo {author} {\bibfnamefont
  {E.}~\bibnamefont {Kozik}}, \bibinfo {author} {\bibfnamefont {X.-W.}\
  \bibnamefont {Liu}}, \bibinfo {author} {\bibfnamefont {A.~J.}\ \bibnamefont
  {Millis}}, \bibinfo {author} {\bibfnamefont {N.~V.}\ \bibnamefont
  {Prokof'ev}}, \bibinfo {author} {\bibfnamefont {M.}~\bibnamefont {Qin}},
  \bibinfo {author} {\bibfnamefont {G.~E.}\ \bibnamefont {Scuseria}}, \bibinfo
  {author} {\bibfnamefont {H.}~\bibnamefont {Shi}}, \bibinfo {author}
  {\bibfnamefont {B.~V.}\ \bibnamefont {Svistunov}}, \bibinfo {author}
  {\bibfnamefont {L.~F.}\ \bibnamefont {Tocchio}}, \bibinfo {author}
  {\bibfnamefont {I.~S.}\ \bibnamefont {Tupitsyn}}, \bibinfo {author}
  {\bibfnamefont {S.~R.}\ \bibnamefont {White}}, \bibinfo {author}
  {\bibfnamefont {S.}~\bibnamefont {Zhang}}, \bibinfo {author} {\bibfnamefont
  {B.-X.}\ \bibnamefont {Zheng}}, \bibinfo {author} {\bibfnamefont
  {Z.}~\bibnamefont {Zhu}}, \ and\ \bibinfo {author} {\bibfnamefont
  {E.}~\bibnamefont {Gull}},\ }\href {\doibase 10.1103/PhysRevX.5.041041}
  {\bibfield  {journal} {\bibinfo  {journal} {Phys. Rev. X}\ }\textbf {\bibinfo
  {volume} {5}},\ \bibinfo {pages} {041041} (\bibinfo {year}
  {2015})}\BibitemShut {NoStop}%
\end{thebibliography}%

\newpage
\clearpage
\appendix
\section{Supplementary Material}
\renewcommand{\thefigure}{S\arabic{figure}}
\setcounter{figure}{0}

\subsection{Hubbard Chain Rotations}
\subsubsection{Sign Free Basis Rotations}
\begin{figure}[H]
\noindent \centering{}\includegraphics[width=0.5\textwidth]{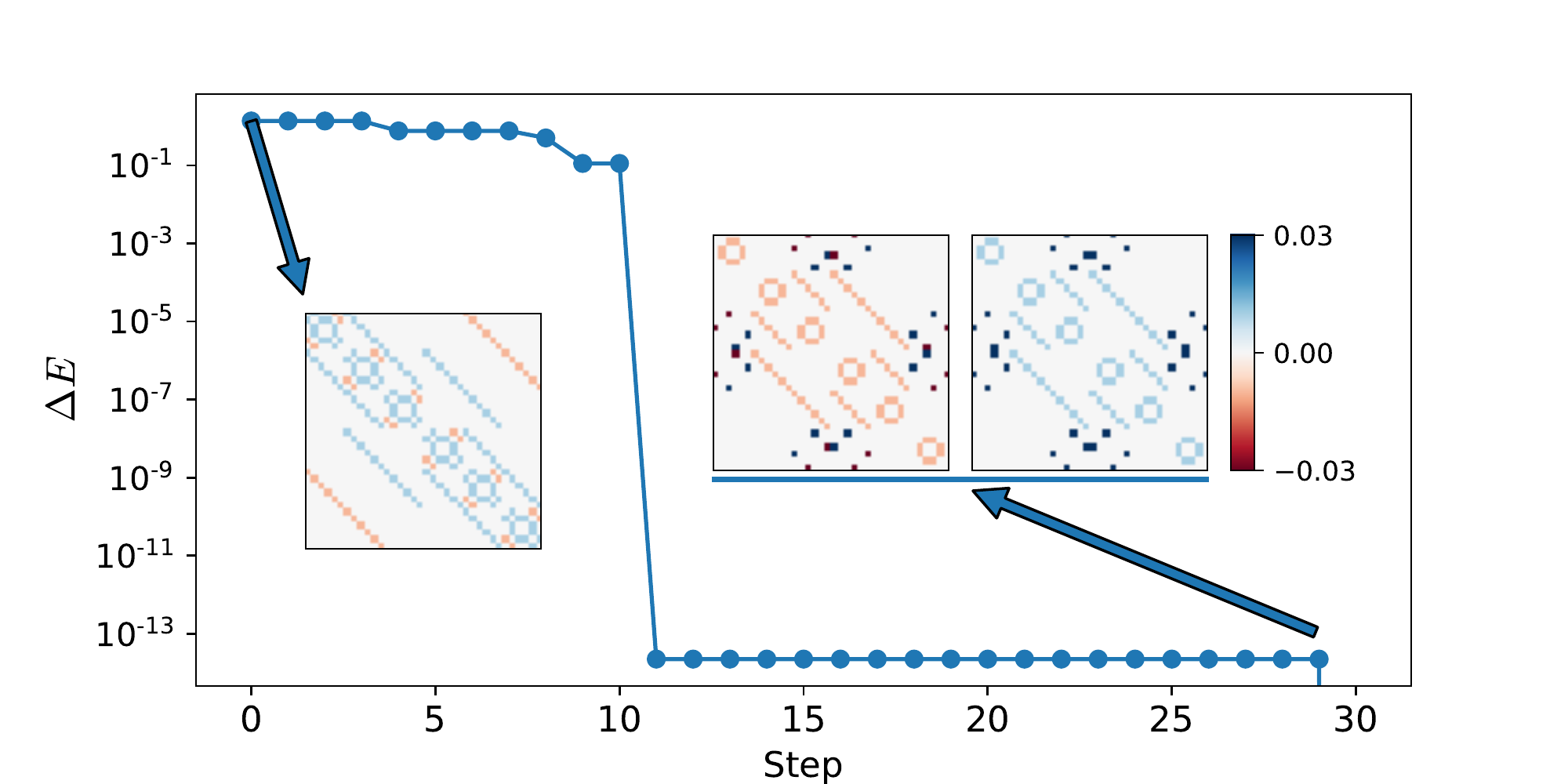}
\caption{$\Delta E$ vs. optimization step for a 4 site Hubbard chain at $n=1$ and $U/t=1$.
\textit{Inset (left)}: Original many-body $G$ with the diagonal removed (corresponding to double occupied states). The red elements correspond to matrix elements which might be inducing a sign problem.
\textit{Inset (middle)}: Corresponding rotated (by eq.~\ref{eq:U_N4}) $G$ with the diagonal removed.
\textit{Inset (right)}: Rotated $G$ with the application of a  `Marshall sign rule' to make the already existing lack of sign problem in the middle inset explicit.}
\label{Fig:rotatedG} 
\end{figure}

Our algorithm developed two basis rotations that completely removed the sign problem for all values of $U$ but did not diagonalize the Hamiltonian directly. For the $N=3$, $n=2/3$ unitary matrix is
\begin{equation}
U= \begin{pmatrix}
-\frac{1}{\sqrt{2}} & 0 & \frac{1}{\sqrt{2}}\\
 \frac{1}{\sqrt{2}} & 0 & \frac{1}{\sqrt{2}}\\
                  0 & 1 & 0
\end{pmatrix},
\label{eq:U_N3}
\end{equation}
and for $N=4$, $n=1$ the unitary matrix is 
\begin{equation}
U = \frac{1}{\sqrt{2}}
\begin{pmatrix}-1 & 0 & 0 & 1\\
                0 & 1 & 1 & 0\\
                1 & 0 & 0 & 1\\
                0 & -1 & 1 & 0
\end{pmatrix}.
\label{eq:U_N4}
\end{equation}
We illustrate the $N=4$ rotation and optimization in fig.~\ref{Fig:rotatedG}, showing the Hamiltonian under the rotation of the above unitary. 

\subsubsection{8 Site Hubbard Model Optimization}

The final unitary for the 8 site Hubbard chain is
\begin{widetext}
\[
U= \begin{pmatrix}
 -0.314786 &  0.125218 & -0.326724 & 0.0175546 &  0.465212 &  0.366267 &  0.415591 &  0.504861 \\
-0.0370841 & -0.251754 & -0.321497 & -0.511366 & 0.00856541 & -0.497735 & -0.322363 &  0.467608 \\
 0.390442 &  -0.48436 & -0.326193 & 0.0390196 & -0.529225 &  0.362701 &  0.258697 &  0.162699 \\
 -0.527768 & -0.447829 & -0.355797 &  0.509121 & 0.0331845 & -0.0526765 & -0.284827 & -0.223855 \\
 0.432137 & -0.141862 & -0.390688 & -0.0378131 &  0.500662 & -0.290055 &  0.312168 & -0.454778 \\
 -0.110209 &  0.243491 & -0.398565 & -0.493372 & -0.047652 &  0.494441 & -0.344209 &  -0.40134 \\
 -0.258416 &  0.472052 & -0.373214 & 0.0537856 & -0.497251 & -0.393736 &  0.394274 & -0.102314 \\
 0.450426 &  0.428415 & -0.325755 &  0.479303 & 0.0461818 & 0.0122356 & -0.451591 &  0.267417 \\
\end{pmatrix}.
\]
\end{widetext}

\subsection{Exploring the $4\times 4$ Hubbard Model}
\subsubsection{Bulk Measurement of the Sign problem}

\begin{figure}[H]
    \noindent \centering{}\includegraphics[width=\columnwidth]{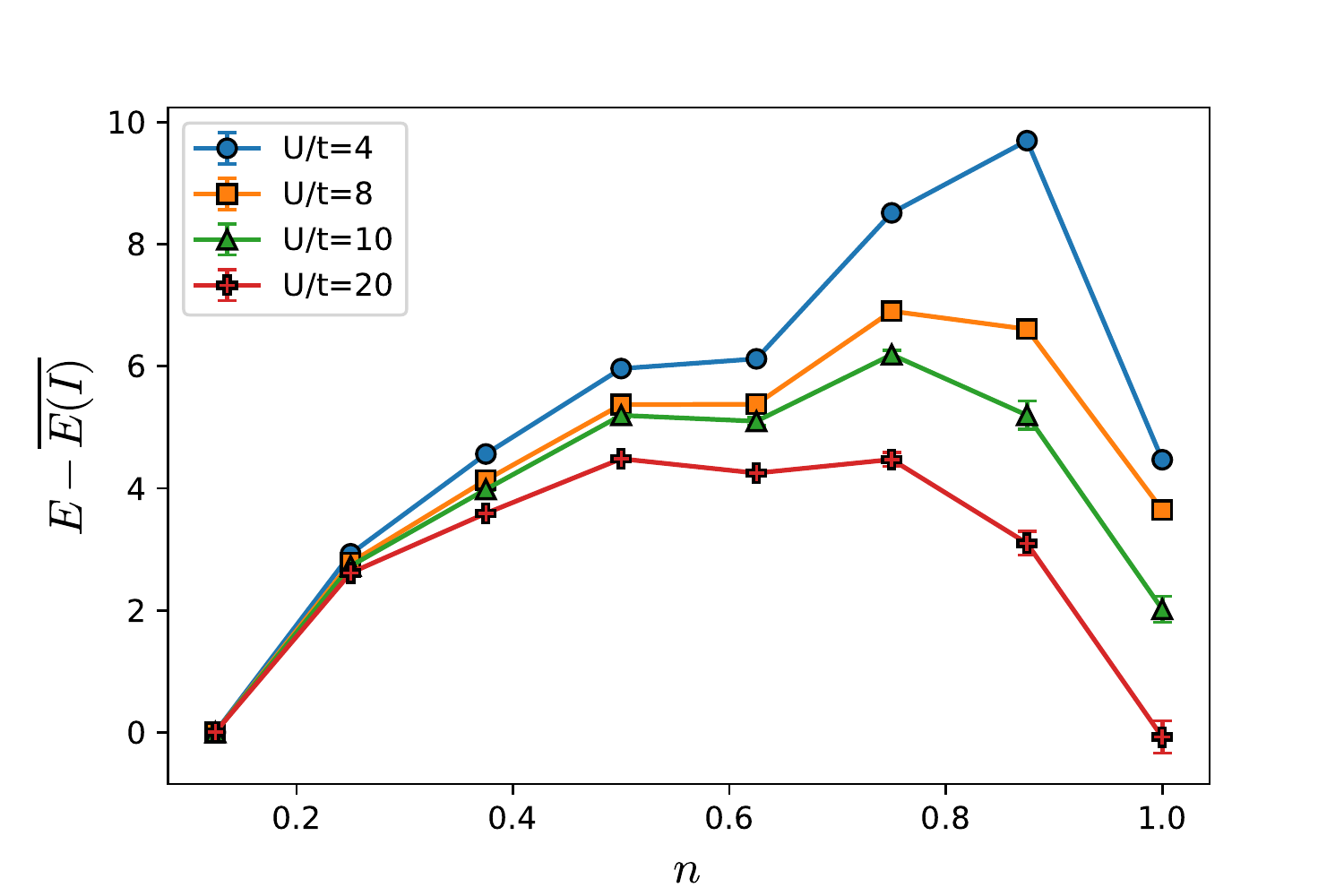}
    \caption{Direct evaluation of the sign problem error $\Delta E$ using PQMC evaluated without a rotation on a 4x4 Hubbard model.}
    \label{Fig:deltaLamVsDope4x4} 
\end{figure}

Using PQMC, we can evaluate the sign problem directly for the 4x4 Hubbard model. We measure $\Delta E = E-\overline{E(I)}$ in fig. \ref{Fig:deltaLamVsDope4x4} using energies from \cite{Dagotto1992} with fixed $S_z=0$. As expected, for intermediate to high $U$ for any number of electrons results in a sign problem. For $n=0.875$, there is a maximum in the sign problem for $U/t=4$, a doping that has been noted for its difficulty in numeric simulations \cite{LeBlanc2015}. Even at half-filling, where other techniques do not have a sign problem \cite{Wang2015}, PQMC simulations would suffer strongly.

\subsubsection{Optimized Unitary Rotation}
\begin{figure}[h]
    \noindent \centering{}\includegraphics[width=\columnwidth]{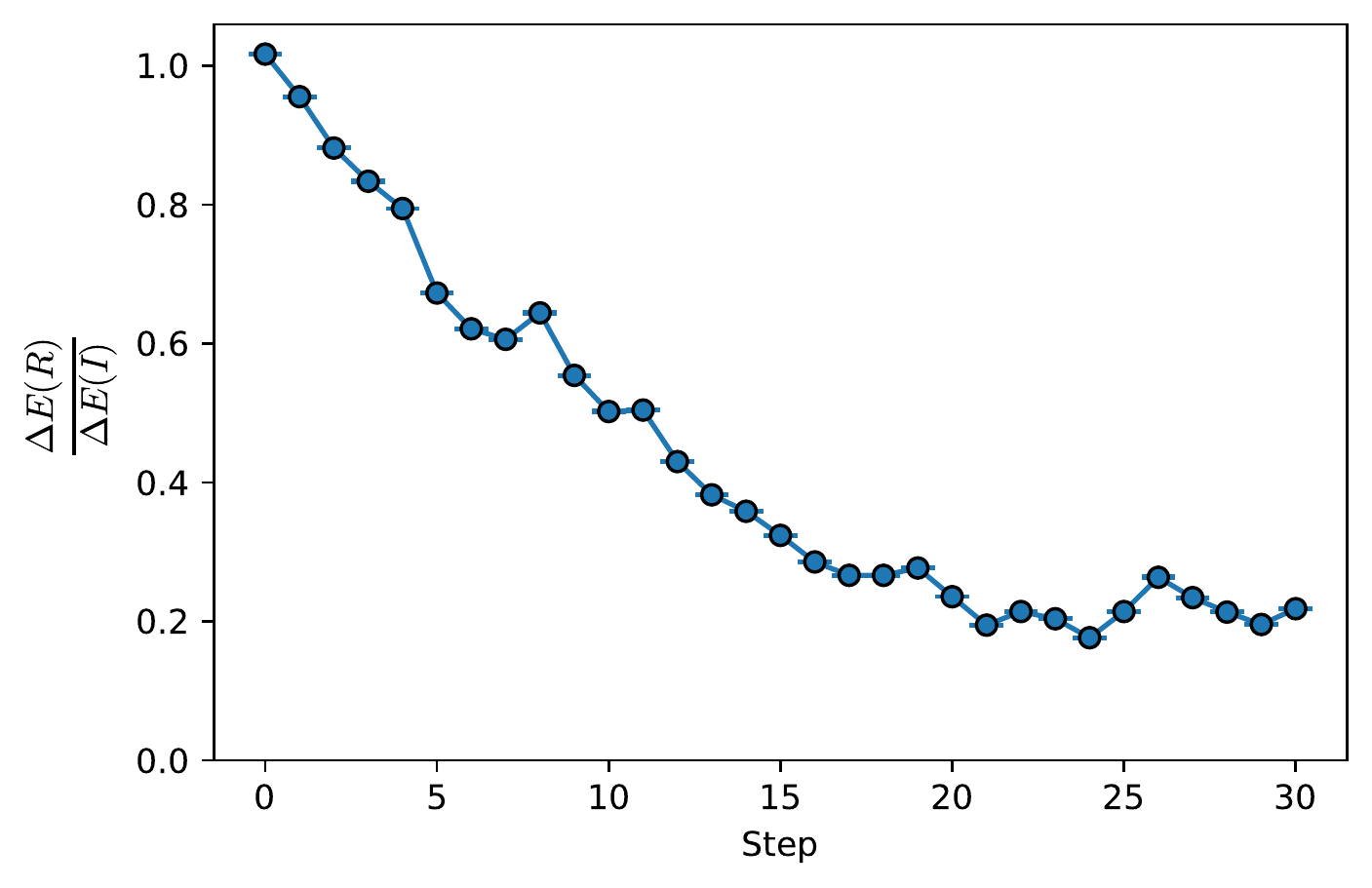}
    \caption{Ratio of $\Delta E$ of the $4\times 4$ optimized unitary to not rotating ($R=I$) during the  optimization in fig.~\ref{Fig:optimization}}
    \label{Fig:deltaEOptmization} 
\end{figure}

During the optimization, shown in fig.~\ref{Fig:optimization}, we can track the ratio of $\Delta E$ to compare with the true ground state fermionic energy known from exact diagonalization. This is tracked in fig.~\ref{Fig:deltaEOptmization}. At the optimal point, step 24, we reach a $\Delta E(R)/\Delta E(I)=0.176(1)$, while the rounded sparse unitary reaches $\Delta E(R)/\Delta E(I)=0.02$, a factor of 8.8 improvement. 

The rounded sparse unitary can described as 
\begin{equation}
    b_{(i,j)}=\sum_{k}U_{jk} \frac{1}{\sqrt{2}}\left(c_{(i+1,k)}- s_{(i-1)} c_{(i-1,k)}\right)
\end{equation}
where 
\begin{equation}
    U_{jk} = \frac{1}{\sqrt{2}} \begin{pmatrix}
    0 & -1 & 0 & -1\\
    1 & 0 & -1 & 0\\
    0 & 1 & 0 & -1\\
    1 & 0 & 1 & 0
    \end{pmatrix}
\end{equation}
and $s_1=s_2=-1$; $s_3=s_4=1$ and the indices are taken $\mod 4$. To create this unitary, we start with the original optimized unitary $R_ij$ at step 24 of our optimization, we then set to zero any values $\left| R_{ij}\right|<0.1$ and equally weight the remaining values, keeping the original signs intact. This new unitary has 64 non-zero entries, out of a possible 256 entries.   

\begin{figure}[h]
    \noindent \centering{\includegraphics[width=\columnwidth]{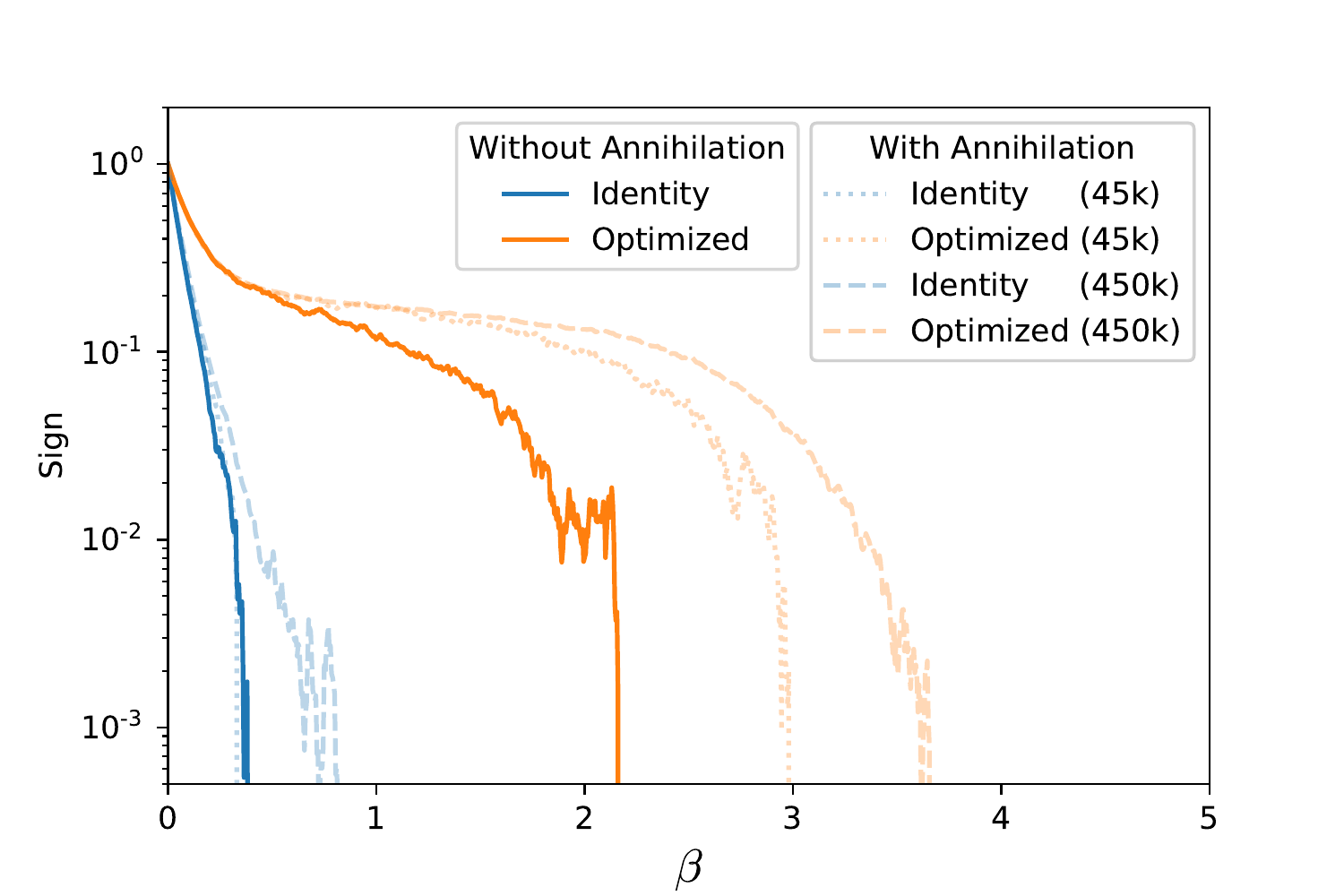}}
    \caption{Average sign of the walkers using a single PQMC simulation for the identity and optimized rotation both with and without annihilation.  We resample/annihilate every $T=0.005$ in PQMC; parentheses in legend denote number of walkers used.}
    \label{Fig:optimizationAnn} 
\end{figure}

In fig~\ref{Fig:optimizationAnn} we compare the average sign of the identity rotation against our optimized basis finding significant improvement and allowing projection out to much larger $\beta$ albeit at the cost of a denser basis.  We also compare against another explicit approach to mitigate sign-problems, the use of annihilation \cite{doi:10.1063/1.3193710,Kalos2000} of opposite signed walkers as is used in FCIQMC (while the annihilation technique of FCIQMC is used here, we do not apply the initiator approximation).  While the sign-problem is insensitive to the number of walkers used in PQMC, the number of walkers does matter when using annihilation.  
Using a number of walkers which is approximately $1\%$ of the Hilbert space size, we find two interesting things.  First, in this example, basis rotation improves the sign problem significantly more then improvements seen using annihilation.  Second, annihilation and basis rotation can be combined and, in this situation, we find annihilation actually helps more in the case where the basis rotation has already partially mitigated the sign problem.

\subsection{Scaling the optimization - $6\times6$ Hubbard Model}
\begin{figure}[h]
    \noindent \centering{}\includegraphics[width=\columnwidth]{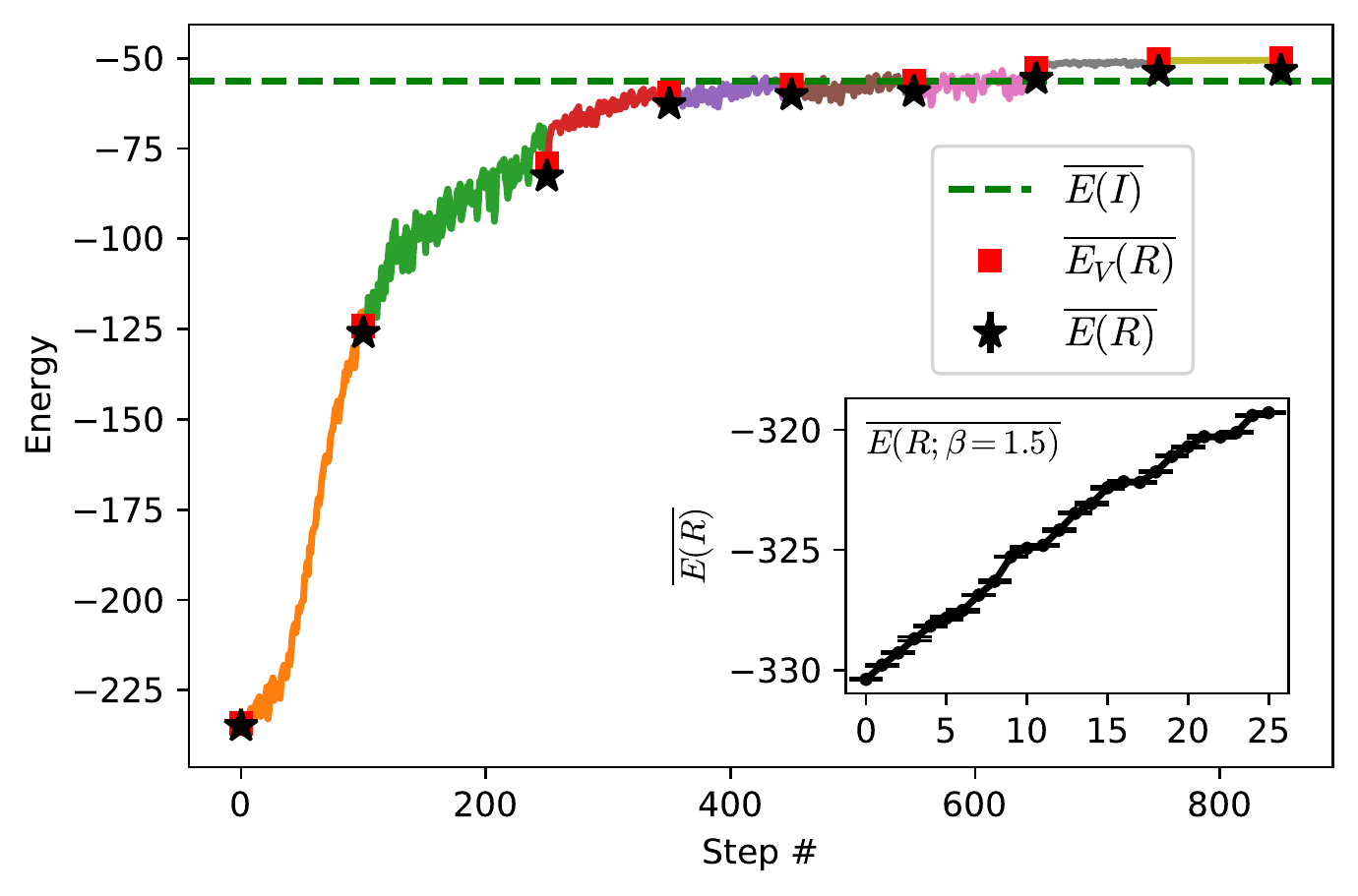}
            
    \caption{Variational optimization of $6 \times 6$ Hubbard model showing both noisy values of  $\overline{E_V(R)}$ used in optimization (colored lines) as well as accurate values of   $\overline{E_V(R)}$ (red squares) and $\overline{E(R)}$ (stars).  Each color indicates a change in hyperparameters of optimization (see supplementary). \textit{Inset:} Projector optimization (using $\overline{E(R;\beta=1.5)}$) from different initial conditions then (b) showing $\overline{E(R)}$.}
    \label{Fig:optimization6x6} 
\end{figure}
Because the scaling or our algorithm with system size is polynomial, one can perform projector optimization of $\overline{E(R)}$ over larger system sizes;  we optimize  the 630-parameter space of a $6 \times 6$ Hubbard model at $U/t=1$ and $n=0.5$ over a random starting basis using finite differences.  We successfully use this method over 25 steps to improve the average sign from a random rotation by $\exp\left[11.09(4)\beta\right]$ (see the inset of fig.~\ref{Fig:optimization6x6}).

Similarly, we can perform optimization on $\overline{E_V(R)}$ using finite differences, shown in fig.~\ref{Fig:optimization6x6}. We update optimization parameters (hyperparameters) every 100 steps or so, corresponding color changes. Steps 0-549 use 11 states to calculate derivatives, steps 550-749 use 21 states, and steps 750-850 use 41 states. The step size $\gamma$ also varies: steps 0-249 use $\gamma=0.01$, steps 250-649 use $\gamma=0.005$, steps 650-749 use $\gamma=0.001$, and steps 750-849 use $\gamma=0.0001$. 
The final rotation found improves the average sign by $\approx \exp\left[2.95(3)\beta\right]$ over not rotating ($R=I$).

\subsection{Hyperparameters and Optimization of the $16\times4$ Optimization}
During optimization of the $16\times 4$ Hubbard model at $n=0.5$ and $U/t=2$, we use 20 walkers fixed throughout the 18800 steps, 19 randomly chosen alongside the completely doubly occupied state. For the first 100 steps we use $\gamma=0.01$ then $\gamma=0.001$ from steps 100-500, followed by  $\gamma=0.0001$ for 500 steps. Afterward, we apply 8 sequences of 100 steps of $\gamma=0.001$ followed by $\gamma=0.0001$ for 1000 steps, until step 8800 which uses $\gamma=10^{-5}$ 

To round the final noisy unitary, we set to zero any values $\left| R_{ij}\right|<0.1$ and equally weight the remaining values, keeping the original signs intact. This new unitary has 160 non-zero entries, out of a possible 4096 entries.   

\subsection{Energy Scaling of the Width-4 Hubbard Models}

\begin{figure*}[th]
    \noindent \centering{}\includegraphics[width=2\columnwidth]{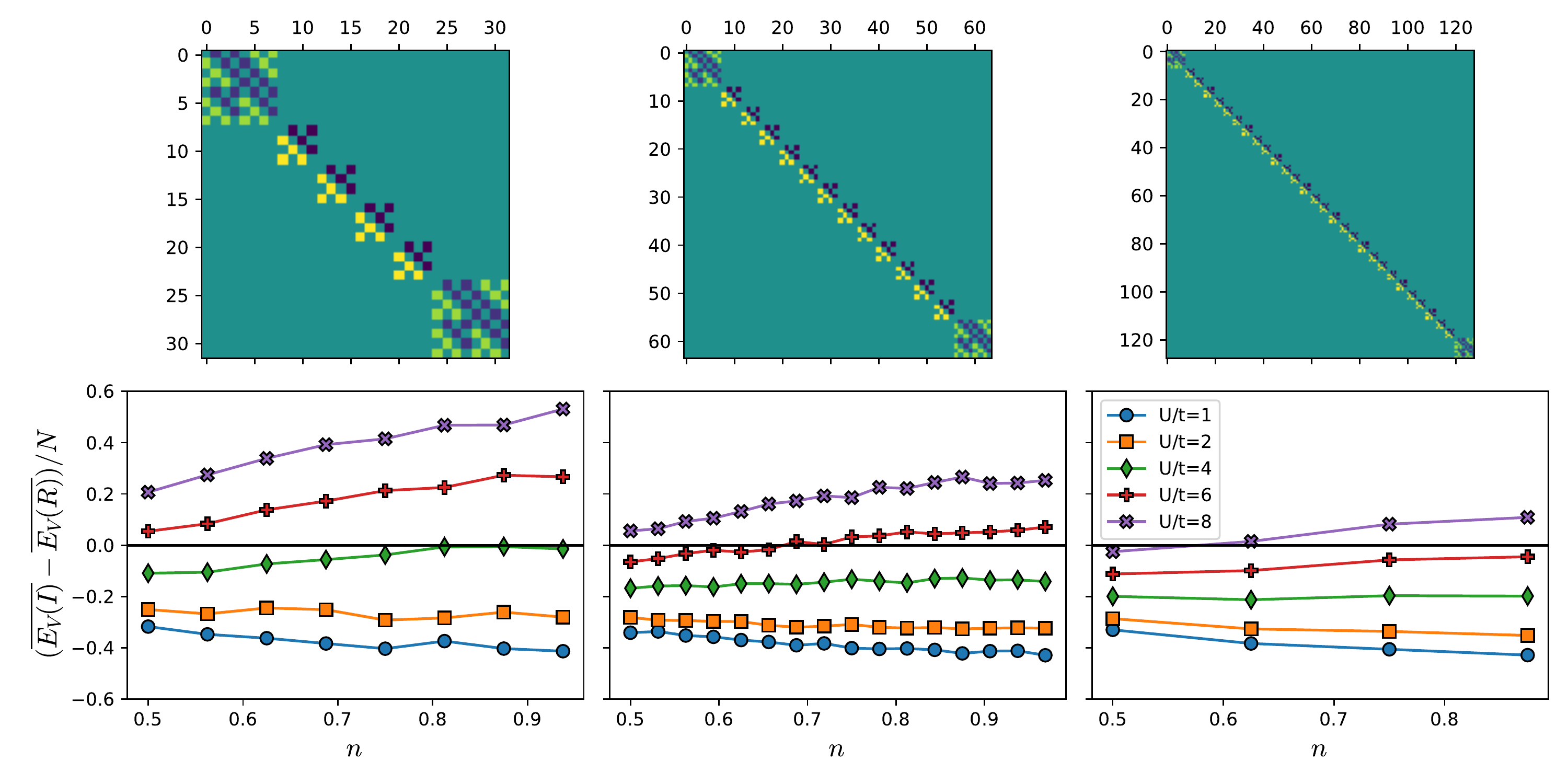}
            
    \caption{\textit{Top Row:} Unitary Rotations used for the $L=8,16,32$ $L\times4$ Hubbard model respectively.
    \textit{Bottom Row:} Energy improvement per site compared to the identity. Negative values represent decreasing the `badness' of the sign problem, positive values represent a worse sign problem than not rotating. }
    \label{Fig:4x_beta0} 
\end{figure*}
First, we measure for various doping $\overline{E_V(R)}$ for the $8\times4$, $16\times4$ and $32\times 4$ lattices using using 500, 500, and 250 sampled states respectively, shown in fig.~\ref{Fig:4x_beta0}. Using the unitary construction defined by eqs.~\ref{eq:colRot},\ref{eq:edgeRot} (shown in the first row of fig.~\ref{Fig:4x_beta0}), as system sizes increase, the larger $U/t$ values begin to have less of a sign problem. For smaller $U/t$, the energy gap per site remains relatively constant, suggesting that the rotation is mitigated compared to rotating for $L\gg1$. 
\begin{figure}[th]
    \noindent \centering{}\includegraphics[width=\columnwidth]{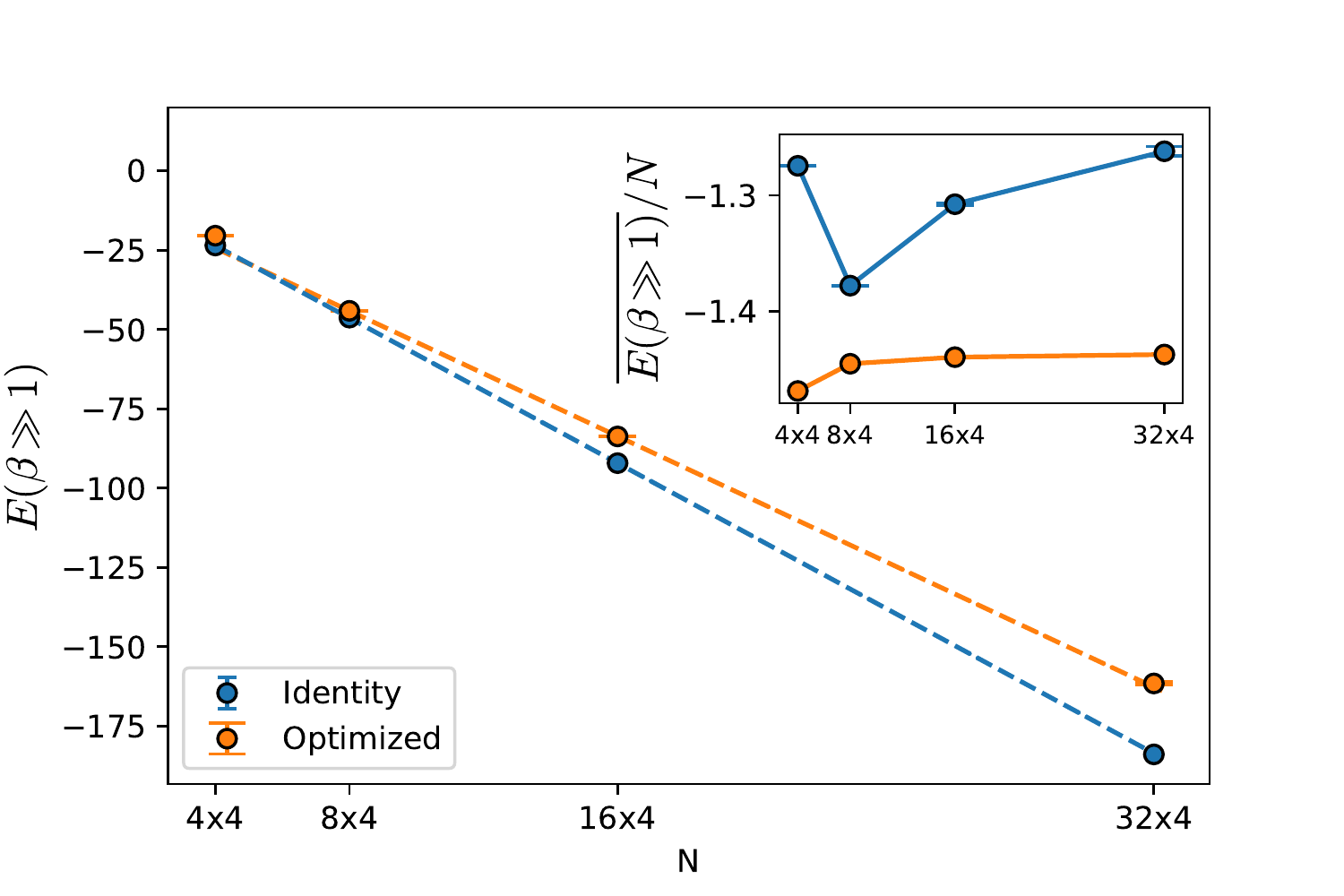}
            
    \caption{Energy $\overline{E(R)}$ at $\beta\gg1$ for the optimized rotation and  for various width-4 Hubbard models. PQMC is done in continuous time with a resampling rate of  $T=0.1$ to a projection time of $\beta=20$.
    \textit{Inset:} Energy per site $\overline{E(R)}/N$ of the same data. }
    \label{Fig:4xthermo} 
\end{figure}

To test this, we look at $\overline{E(I)}-\overline{E(R)}$ at $U/t=4$ and $n=0.5$ for $L=4,8,16,32$ $L\times 4$ Hubbard models, shown in fig.~\ref{Fig:4xthermo}. Fitting the energies to a line (excluding $L=4$) we find a new a variational lower bound to the ground state energy of $-1.23544(5)$/site, greater than the lower bound of -1.429490(4)/site from not rotating.

\end{document}